\documentclass[%
 aip,
 amsmath,amssymb,
 reprint,%
]{revtex4-1}

\usepackage{graphicx}% Include figure files
\usepackage{dcolumn}% Align table columns on decimal point
\usepackage{bm}% bold math
\usepackage{amsmath,amssymb}
\usepackage[utf8]{inputenc}
\usepackage[T1]{fontenc}
\usepackage{mathptmx}
\usepackage{etoolbox}

\makeatletter
\def\@email#1#2{%
 \endgroup
 \patchcmd{\titleblock@produce}
  {\frontmatter@RRAPformat}
  {\frontmatter@RRAPformat{\produce@RRAP{*#1\href{mailto:#2}{#2}}}\frontmatter@RRAPformat}
  {}{}
}%
\makeatother
\begin{document}

\preprint{AIP/123-QED}

\title[Sample title]{Sample Title:\\with Forced Linebreak}
% Force line breaks with \\
%\title[]{Dynamics of a classical, active wave-particle entity with infinite wave-memory driven by the diffusionless Lorenz equations}
\title[]{Infinite-memory classical wave-particle entities, attractor-driven active particles and the diffusionless Lorenz equations}
%\title[]{A classical, active wave-particle entity with infinite wave-memory: using phase-space attractors of diffusionless Lorenz equations to of pilo-wave hydrodynamics in the high-memory regime}
% Force line breaks with \\
\author{R. N. Valani}
 %\altaffiliation[Also at ]{School   of   Mathematical   Sciences,    University   of   Adelaide,    South   Australia,    Australia}%Lines break automatically or can be forced with \\
%\author{B. Author}%
 \email{rahil.valani@adelaide.edu.au}
\affiliation{ 
School of Computer and Mathematical Sciences, University of Adelaide, South Australia 5005, Australia
}%

\date{\today}% It is always \today, today,
             %  but any date may be explicitly specified

\begin{abstract}
A classical wave-particle entity (WPE) can materialize as a millimeter-sized droplet walking horizontally on the free surface of a vertically vibrating liquid bath. This WPE comprises a particle (droplet) that shapes its environment by locally exciting decaying standing waves, which in turn guides the particle motion. At high amplitude of bath vibrations, the particle-generated waves decay very slowly in time and the particle motion is influenced by the history of waves along its trajectory. In this high-memory regime, WPEs exhibit hydrodynamic quantum analogs where quantum-like statistics arise from underlying chaotic dynamics. Exploration of WPE dynamics in the very high-memory regime requires solving an integro-differential equation of motion. By using an idealized one-dimensional WPE model where the particle generates sinusoidal waves, we show that in the limit of infinite memory, the system dynamics reduce to a $3$D nonlinear system of ordinary differential equations (ODEs) known as the diffusionless Lorenz equations (DLEs). We use our algebraically simple ODE system to explore in detail, theoretically and numerically, the rich set of periodic and chaotic dynamical behaviors exhibited by the WPE in the parameter space. Specifically, we link the geometry and dynamics in phase-space of the DLE system to the dynamical and statistical features of WPE motion, paving a way to understand hydrodynamic quantum analogs using phase-space attractors. Our system also provides an alternate interpretation of an attractor-driven particle, i.e.~an active particle driven by internal state-space variables of the DLE system. Hence, our results might also provide new insights in modeling active particle locomotion.

\end{abstract}

\maketitle

\begin{quotation}
Walking and superwalking droplets can emerge on the free surface of a vibrating liquid bath and constitute a classical, active wave-particle entity (WPE). The particle (droplet) repeatedly generates decaying localized standing waves which in turn guide the motion of the particle. In the high-memory regime, the particle-generated waves decay very slowly in time and the WPE motion is affected by the waves generated in the distant past along the particle's trajectory. Chaotic dynamics of the WPE in this regime can result in wave-like statistics and the WPE displays hydrodynamic quantum analogs. We use an idealized model of a one-dimensional WPE to investigate this very high-memory regime and show that the dynamics of the system reduce to one of the algebraically simplest dynamical system that exhibits chaos - the diffusionless Lorenz equations (DLEs). The system also forms an example of an attractor-driven particle, i.e.~an active particle driven by the DLE residing in its internal state-space. By investigating this system in detail, we link the phase-space geometry, dynamics and bifurcations of the DLE system to the motion and trajectories of the particle, giving us new insights into using phase-space attractors to understand hydrodynamic quantum analogs and model active particle motion.
\end{quotation}

\section{\label{sec: intro} Introduction}

%\begin{itemize}
%    \item Start by introducing walkers. Maybe its a good idea to introduce in the context of active particle with self-generated wave forms and then maybe I can introduce walkers as a particular case of that
%    \item Mention about the studies of walkers done at high memory and how they have been approximated as random walks
%    \item Mention about the studies of simplified models of walkers and mention the transformation from integrodifferential equation to LOrenz system
%    \item Mention in this paper we study various aspects of infinite memory walker in 1D and the corresponding dynamical system
%    \item Mention the different simple systems that generate chaotic flows and where they are applied
%\end{itemize}

Active particles are self-propelled entities that extract energy from their surroundings and convert it into directed motion. Active entities can be found at all scales in nature, for example, macroscopic living organisms such as humans, birds and fish or microorganisms such as sperm cells, bacteria and algae. They also arise in artificial systems such as active colloidal particles~\citep{PhysRevLett.99.048102} and microrobots~\cite{Palagi2018}. In some active particle systems, the particle motion is guided by interaction with an environment that is itself created by the particle. For example the motion of autophoretic microswimmers is powered by chemical activity at the particle's surface which generates long-lived chemical gradients~\citep{10.1063/1.4810749,stopgoswim,kailasham_khair_2022}. This self-generated chemical environment in turn guides the motion of the microswimmer. A curious hydrodynamic system of active entities that are driven by self-generated dynamic environments is walking and superwalking droplets~\cite{Couder2005WalkingDroplets,superwalker}. In this system, millimeter-sized droplets of oil walk horizontally while periodically bouncing on the free surface of a vertically vibrating bath of the same liquid. Each bounce of the droplet excites a spatially localized standing wave that slowly decays in time, on the liquid surface. The droplet then interacts with these waves on subsequent bounces to propel itself horizontally, giving rise to a classical, active wave-particle entity (WPE) on the liquid surface.

In the high-memory regime of WPEs, the waves generated by the droplet on each impact decay very slowly in time, hence the droplet's walking dynamics are not only influenced by the recent waves generated by the droplet, but also by the waves generated in the distant past along its trajectory. This gives rise to \emph{path memory} in the system and make the dynamics non-Markovian. Remarkably, in the high-memory regime, WPEs have been shown to exhibit hydrodynamic analogs of various quantum systems~\citep{Bush2018,Bush_2020}. 

In the absence of obstacles and other droplets, a WPE typically move steadily along a straight line. However, in the high-memory regime, it has been observed in experiments that this steady motion of WPEs can become unstable and one observes speed oscillations~\citep{Bacot2019}. To capture the experimentally observed walking dynamics of a WPE, many theoretical models have been developed over the years~\citep{Rahman2020review,Turton2018}. One such routinely used model with intermediate complexity is the stroboscopic model of \citet{Oza2013}. This model provides a trajectory equation for the two-dimensional horizontal walking dynamics in the form of an integro-differential equation of motion. With the aid of these models, WPE motion can be explored in the very high-memory regime that are currently not achievable in experiments. Simulations of an individual WPE in this regime have shown the emergence of rich dynamical behaviors such as a run-and-tumble-like diffusive motion~\citep{Hubert2019,durey2021}.

%To capture the experimentally observed walking dynamics of WPEs, many theoretical models have been developed over the years~\citep{Rahman2020review}. A routinely used model with intermediate complexity is the stroboscopic model of \citet{Oza2013}. It provides a trajectory equation for the horizontal walking dynamics in the form of an integrodifferential equation of motion.

To explore the dynamics of WPEs and their hydrodynamic quantum analogs beyond the restricted parameter space of experiments,~\citet{Bush2015} proposed the framework of generalized pilot-wave dynamics. It is a theoretical abstraction rooted in the walking-droplet system that has allowed for exploration of a broader class of dynamical systems and the discovery of new quantum analogs~\citep{Bush_2020}. The generalized pilot-wave framework has motivated exploration of idealized pilot-wave systems that consider dynamics of WPEs in one horizontal dimension~\citep{phdthesismolacek,Durey2020,Durey2020lorenz,ValaniUnsteady,Valani2022ANM,Perks2023,Rahman2018,Gilet2014}. Typically in the stroboscopic model of \citet{Oza2013}, one uses a Bessel function wave form to reasonably capture the experimentally observed waves generated by the droplet. This wave form has two key features: spatial oscillations and spatial decay. However, choosing a simple sinusoidal wave form that only captures spatial oscillations for a single one-dimensional WPE, results in reduction of the infinite-dimensional dynamical system generated by the integro-differential equation~\citep{Oza2013,Durey2020} to a low dimensional system of ODEs that can be mapped onto the classic Lorenz system~\citep{Durey2020lorenz,ValaniUnsteady}. Moreover, \citet{Valanilorenz2022} showed a general transformation for a $1$D WPE that can map the infinite-dimensional integro-differential equation to low-dimensional Lorenz-like equations for certain choices of wave forms. 

One of the key features associated with many of the hydrodynamic quantum analogs of WPEs, is the emergence of wave-like statistics from underlying chaotic dynamics in the high-memory regime~\citep{Saenz2017,PhysRevE.88.011001,Giletconfined2016,Cristea,durey_milewski_wang_2020,Durey2020}. However, the non-Markovian nature of the system in this regime makes the integro-differential dynamical equation analytically intractable, hence making it difficult to comprehensively explore the underlying chaos that gives rise to emergent wave-like statistics. Motivated by this, in this paper, we consider the infinite-memory limit in the Lorenz model of a $1$D WPE that employs a sinusoidal wave form~\citep{Valanilorenz2022}. Contrary to intuition, we show that this infinite-memory limit reduce the system dynamics to one of the algebraically simplest chaotic systems, the diffusionless Lorenz equations (DLE)~\citep{difflesslorenz2000}. We explore this single-parameter dynamical system in detail and connect the geometry, dynamics and bifurcations of phase-space attractors to the dynamical and statistical features of WPE motion. 

As we will show at the end of Sec.~\ref{sec: DS}, our infinite-memory WPE dynamical system may also be interpreted as an attractor-driven active particle~\cite{Valaniattractormatter2023}, i.e.~an overdamped active particle driven by internal low-dimensional chaotic DLE system, as opposed to constant self-propulsion and stochastic noise that is generally considered in modeling traditional active particles~\cite{Shaebani2020,Romanczuk2012}. In addition to the WPE system, there are examples of active particles in nature where signatures of low-dimensional chaos have been observed in motility of organisms. Examples include movement patterns of ants~\citep{cole1991}, mud snails~\citep{Reynolds2016}, amoebas~\cite{Miyoshi2001} and worms~\cite{Ahamed2021,Loveless2021}. In light of these examples, our dynamical system may also be viewed as describing the dynamics of a simple active particle driven by internal complexity which is modeled by a low-dimensional chaotic system of DLE~\citep{Valaniattractormatter2023}.
%\textbf{Mention here that the dynamics of the WPE at high but finite memory has been explored in various papers. Including \citet{Hubert2022}.}

The paper is organized as follows. In Sec.~\ref{sec: DS} we consider the stroboscopic model of \citet{Oza2013} and derive the diffusionless Lorenz equations (DLE) that governs the dynamics of our system, In Sec.~\ref{sec: lin stab} we consider the steady states of the particle goverened by DLE and determine their stability by performing linear stability analysis. In Sec.~\ref{sec: PS space} we numerically explore the unsteady dynamical states in the parameter-space of the dynamical system linking the phase-space features of the DLE system with particle's dynamical and statistical features. We conclude in Sec.~\ref{sec: conclusion}.

%\section{\label{sec: DS} Wave-particle entity with infinite memory}
\section{\label{sec: DS} Deriving the dynamical system}

\begin{figure}
\centering
\includegraphics[width=\columnwidth]{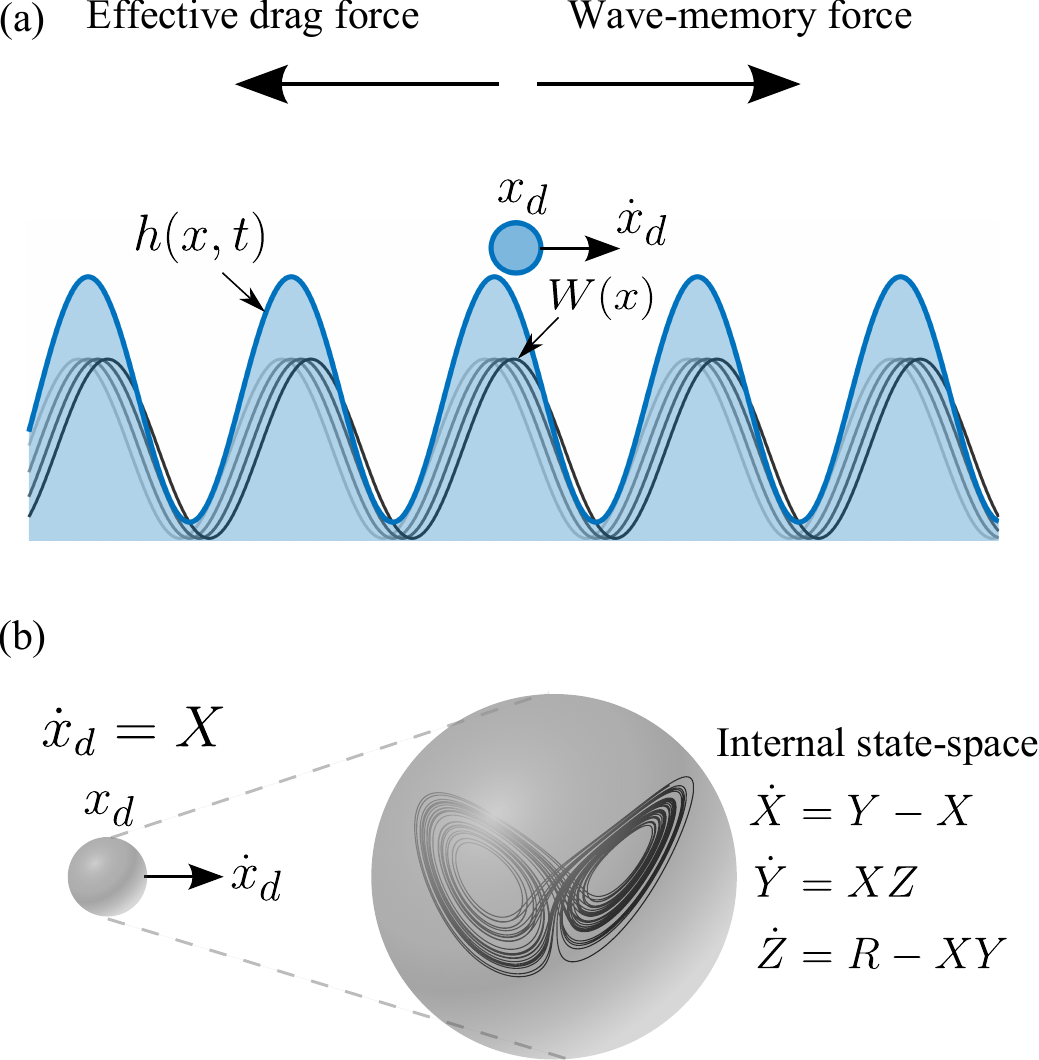}
\caption{Schematic illustrating two different viewpoints of our system as an infinite-memory wave-particle entity (WPE) and an attractor-driven particle. (a) Schematic of the one-dimensional WPE. A particle located at $x_d$ and moving horizontally with velocity $\dot{x}_d$ generates, at each instant, a wave with spatial form $W(x)=\cos(x)$ centered at its position $x_d$ (black and gray curves, with the higher intensity of the color indicating the waves created more recently). Superposition of the individual waves generated by the particle continuously along its trajectory results in the wave field $h(x,t)$ (blue filled area). The dynamics of the particle are governed by two forces: (i) a propulsive wave-memory force proportional to the gradient of its self-generated wave field $h(x,t)$, and (ii) an effective drag force proportional to its velocity. Thus, the wave and the particle are coupled to each other forming a WPE. (b) An alternate interpretation of the system as an one-dimensional attractor-driven particle, i.e.~an active particle who motion is driven by internal low-dimensional chaos arising from DLE. The three-dimensional internal state-space formed by DLE with variables $(X,Y,Z)$ drives particle motion via the overdamped equation of motion $\dot{x}_d=X$.}
\label{Fig: schematic}
\end{figure}

%\begin{itemize}
%    \item Start with dimensional integrodifferential equation of motion
%    \item Show the correspondance with LOrenz equation for a given choice of non-dimensionlozation
%    \item Show that for a different choice of dimensionless variable, we get memory as a explicit parameter. And show that in the limit that this goes to infinity, we get the simplified Lorenz system
%    \item Mention how these equation arises in different studies and how it has been shown to be one of the simpliest chaotic system
%\end{itemize}

Consider a droplet (particle) bouncing periodically on a vertically vibrating bath of the same liquid while moving horizontally in two dimensions. Using the fact that the time scale of vertical bouncing is very small compared to horizontal walking, \citet{Oza2013} developed a theoretical stroboscopic model that averages over the vertical periodic bouncing motion of the particle and provides a continuum description of the horizontal walking motion. Let the particle be located at horizontal position $\mathbf{x}_d$ and moving with horizontal velocity $\mathbf{\dot{x}}_d$ while continuously generating axisymmetric standing waves that are centered at the particle location, have spatial structure $W(\mathbf{|x|})$ and decay exponentially in time. This results in the following equation of motion for the horizontal dynamics of the WPE~\citep{Oza2013}
\begin{equation}\label{eq: dim1}
    m \ddot{\mathbf{x}}_d + D \dot{\mathbf{x}}_d = - m g \nabla h(\mathbf{x}_d,t),
\end{equation}
where an overdot denotes a time derivative. The left-hand-side of Eq.~\eqref{eq: dim1} is composed of an inertial term $m \ddot{\mathbf{x}}_d$ where $m$ is the droplet mass, and an effective drag force term $D \dot{\mathbf{x}}_d$ where $D$ is a non-negative constant that denotes an effective time-averaged drag coefficient. The term on the right-hand-side of the equation captures the forcing on the particle from its self-generated wave field $h(\mathbf{x},t)$  where $g$ is the gravitational acceleration. This force is proportional to the gradient of the self-generated wave field, $\nabla h(\mathbf{x}_d,t)$, at the particle location. The wave field $h(\mathbf{x},t)$ is calculated through integration of the individual wave forms $W(\mathbf{|x|})$ that are continuously generated by the particle along its trajectory and decay exponentially in time, giving
\begin{equation}\label{eq: dim2}
h(\mathbf{x},t)=\frac{A}{T_F} \int_{-\infty}^{t} W\left( k_F |\mathbf{x}-\mathbf{x}_d(s)| \right)\,\text{e}^{-\frac{(t-s)}{T_F \text{Me}}}\,\text{d}s.
\end{equation}
Here, $k_F=2\pi/\lambda_F$ is the Faraday wavenumber with $\lambda_F$ the Faraday wavelength (i.e. the wavelength of droplet-generated waves), $A$ is the amplitude of surface waves, $\text{Me}$ is the memory parameter that prescribes the decay rate of droplet-generated waves and $T_F$ is the Faraday period (i.e. the period of droplet-generated standing waves). We refer the interested reader to \citet{Oza2013} for more details and explicit expressions for these parameters. 

Hence, Eq.~\eqref{eq: dim1} describes the motion of particle $\mathbf{x}_d(t)$ guided by its underlying wave field $h(\mathbf{x},t)$ which in turn in determined by the history of the particle, $\mathbf{x}_d(s)$ for $s<t$, as per Eq.~\eqref{eq: dim2}. Therefore, Eqs.~\eqref{eq: dim1} and \eqref{eq: dim2} in combination describe the motion of the WPE. Substituting Eq.~\eqref{eq: dim2} in \eqref{eq: dim1}, one gets the following integro-differential equation of horizontal motion~\citep{Oza2013}
\begin{align}\label{eq: dimensional 2D}
    & m \ddot{\mathbf{x}}_d +  D \dot{\mathbf{x}}_d = \\ \nonumber
    \quad
    &\frac{mg A k_F}{T_F} \int_{-\infty}^{t} f\left( k_F |\mathbf{x}_d(t)-\mathbf{x}_d(s)| \right)\frac{\mathbf{x}_d(t)-\mathbf{x}_d(s)}{|\mathbf{x}_d(t)-\mathbf{x}_d(s)|}\,\text{e}^{-\frac{(t-s)}{T_F \text{Me}}}\,\text{d}s.
\end{align}
%The left-hand-side of Eq.~\eqref{eq: dimensional 2D} is composed of an inertial term $m \ddot{\mathbf{x}}_d$ and an effective drag force term $D \dot{\mathbf{x}}_d$, where the overdot denotes a time derivative. The term on the right-hand-side of the equation captures the forcing on the droplet due to its self-generated wave field. As described in Eq.~\eqref{eq: dim2}, the total wave field $h(\mathbf{x},t)$ is calculated through integration of the individual wave forms $W(\mathbf{x})$ that are continuously generated by the particle along its trajectory and decay exponentially in time. The function $f(x)=-W'(x)$ is the negative  gradient of the wave form created by the particle. This wave force is proportional to the gradient of the underlying wave field at the location of the droplet. 
%The parameters in Eq.~\eqref{eq: dimensional 2D} are as follows: $k_F=2\pi/\lambda_F$ is the Faraday wavenumber with $\lambda_F$ the Faraday wavelength, $A$ is the amplitude of surface waves, $\text{Me}$ is the memory parameter that prescribes the decay rate of droplet-generated waves and $T_F$ is the Faraday period (i.e. the period of droplet-generated standing waves). We refer the interested reader to \citet{Oza2013} for more details and explicit expressions for these parameters. 
where $f(\cdot)$ is the negative gradient of the wave form $W(\cdot)$. We start by non-dimensionalizing Eq.~\eqref{eq: dimensional 2D} using $\mathbf{x}'=k_F \mathbf{x}$ and $t'=D t/m$ and dropping the primes on the dimensionless variables results in the following equation
%\begin{align}\label{eq: dimensionaless 1 2D}
%    \kappa \ddot{\mathbf{x}}_d + &  \dot{\mathbf{x}}_d = \\ \nonumber
%    &\beta \int_{0}^{t} f\left( |\mathbf{x}_d(t)-\mathbf{x}_d(s)| \right)\frac{\mathbf{x}_d(t)-\mathbf{x}_d(s)}{|\mathbf{x}_d(t)-\mathbf{x}_d(s)|}\,\text{e}^{-\frac{(t-s)}{\text{Me}}}\,\text{d}s.
%\end{align}
%If we further rescale this equation using $t'=t/\kappa$ and drop the primes then we get
\begin{align}\label{eq: dimensionaless 2 2D}
    \ddot{\mathbf{x}}_d + \dot{\mathbf{x}}_d = 
    R \int_{-\infty}^{t} f\left( |\mathbf{x}_d(t)-\mathbf{x}_d(s)| \right)\frac{\mathbf{x}_d(t)-\mathbf{x}_d(s)}{|\mathbf{x}_d(t)-\mathbf{x}_d(s)|}\,\text{e}^{-\frac{(t-s)}{{\tau}}}\,\text{d}s,
\end{align}
where we have introduced the following non-negative dimensionless parameters~\footnote{The parameter $R$ and $\tau$ are related to $\kappa$ and $\beta$ from \citet{Oza2013} by $R=\beta \kappa^2$ and $\tau=1/\kappa$. However, we note that the time has been scaled differently in both representations.}: a dimensionless wave amplitude $R= m^3 g A k_F^2/(D^3 T_F)$ and a dimensionless memory time $\tau=D T_F \text{Me}/m$. 

A reduction of this model to describe the dynamics of a WPE with only one horizontal dimension is given by the following dimensionless integro-differential equation of motion:~\citep{phdthesismolacek,Durey2020,ValaniUnsteady} 
\begin{align}\label{eq: dimless eq2}
    \ddot{{x}}_d +  \dot{{x}}_d = R \int_{-\infty}^{t} f\left( {x}_d(t)-{x}_d(s) \right)\,\text{e}^{-\frac{(t-s)}{{\tau}}}\,\text{d}s.
\end{align}
%
%In the limit of infinite memory i.e. $\textit{Me}\rightarrow \infty$, this equation reduces to
%\begin{align}\label{Eq: dimless eq2}
%\ddot{x}_d+\dot{x}_d=R\int_{0}^t f(x_d(t)-x_d(s))\,\text{d}s
%\\ \nonumber
%\end{align}
%
\citet{Valanilorenz2022} showed that the the integro-differential equation in \eqref{eq: dimless eq2} can be transformed into an infinite set of coupled ODEs as follows:~\footnote{a different non-dimensionlization was used by \citet{Valanilorenz2022} but the transformation still holds}
\begin{align}\label{eq: gen 1D odes}
    \dot{x}_d&= v\\ \nonumber
    \dot{v}&=M_0 - v\\ \nonumber
    \dot{M}_n &= R f^{(n)}(0) - \frac{1}{\tau} M_n + v M_{n+1}\\ \nonumber
\end{align}
where,
\begin{equation}
    M_n = R \int_{-\infty}^{t} f^{(n)}\left( x_d(t) - x_d(s) \right)\,\text{e}^{-\frac{(t-s)}{{\tau}}}\,\text{d}s,
\end{equation}
and $f^{(n)}(\cdot)$ is the $n$th derivative of the function with respect to its argument.

In experiments, the spatial form of the individual waves generated by a WPE are captured reasonably well by, $W(x)=\text{J}_0(x)$, where $\text{J}_0(x)$ is the Bessel function of the first kind and zeroth order~\citep{Oza2013,Molacek2013DropsTheory}. Sometimes, a spatially decaying exponential envelope is included to further improve the comparison with the experimentally observed wave form~\citep{Turton2018}. There are two key features of the droplet-generated individual waves: (i) spatial oscillations and (ii) spatial decay. \citet{ValaniUnsteady} show that oscillations play a key role in capturing the instability of the steady walking state and such instability can be qualitatively captured using a simple sinusoidal particle-generated wave form such that $W(x)=\cos(x)$ and $f(x)=\sin(x)$. The principal advantage of this simple wave form is that it allows us to transform the system of infinite ODEs in \eqref{eq: gen 1D odes} to a low dimensional Lorenz-like system by~\cite{Durey2020lorenz,Valanilorenz2022} 
\begin{equation}
    \label{lorenz}
    \begin{split}
    \dot{x}_d&=X, \\
    \dot{X}&=Y-X, \\
    \dot{Y}&=-\frac{Y}{\tau}+XZ, \\
    \dot{Z}&=R-\frac{Z}{\tau}-XY,
  \end{split}
\end{equation}
 %This system of equations has only one equilibrium solution corresponding to the wave-particle at a constant position, $x_d=\frac{k\pi}{B}$, $X_0=0$, $Y_0=0$ and $Z_0=0$. We note that when there is no added force in the system there is another equilibrium solution corresponding to a free-walking droplet. The solution is $x_d=\frac{k\pi}{B}$, $X_0=\sqrt{r-1}$, $Y_0=\sqrt{r-1}$ and $Z_0=r-1$. Although not an equilibrium solution to (\ref{lorenz}) the paper will still explore the behavior of the particle with initial walking velocity $X_0=\sqrt{r-1}$. 
where
\begin{align}\label{eq: Y Z}
    Y(t)&=R \int_{-\infty}^{t} \sin\left( x_d(t) - x_d(s) \right)\,\text{e}^{-\frac{(t-s)}{{\tau}}}\,\text{d}s,\\ \nonumber
    Z(t)&=R \int_{-\infty}^{t} \cos\left( x_d(t) - x_d(s) \right)\,\text{e}^{-\frac{(t-s)}{{\tau}}}\,\text{d}s.
\end{align}
 Here, $X=\dot{x}_d$ is the droplet's velocity, $Y=R \int_{-\infty}^{t} \sin(x_d(t)-x_d(s))\,\text{d}s$ is the wave-memory force on the particle (proportional to the gradient of the waves at the droplet's location) 
 and $Z=R\int_{-\infty}^{t} \cos(x_d(t)-x_d(s))\,\text{d}s$ is a dimensionless wave height at the particle location. Note that when solving the system in \eqref{lorenz}, one should be careful in that the initial conditions for $Y(0)$ and $Z(0)$ depend on the particle's history, i.e.~$x_d(s)$ for $s<t$. Moreover, note that our choice of non-dimensionalization results in the wavelength of the particle-generated sinusoidal waves to be $2\pi$. %Although this sinusoidal wave-form may not capture the subtle effects associated with phase-shifts and spatial decay of the Bessel wave form, our focus for this paper is in exploring the rich dynamics of this simplified model that reduced to a simple system of nonlinear ODEs. 

\begin{figure*}
\centering
\includegraphics[width=2\columnwidth]{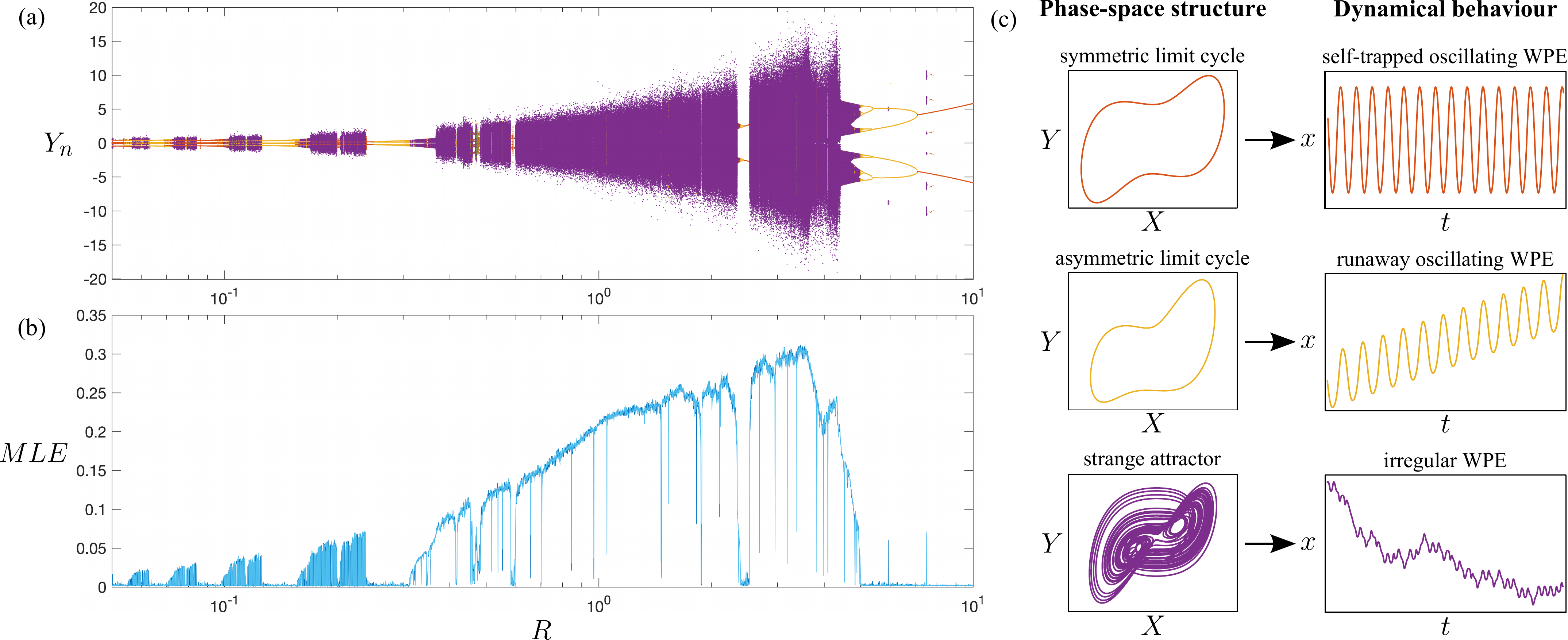}
\caption{Dynamics of the WPE/attractor-driven particle as a function of the dimensionless wave-amplitude parameter $R$. (a) Bifurcation diagram as a function of $R$ showing the wave-memory force $Y_n$ acting on the WPE when its instantaneous velocity is zero i.e.~$X=0$. Trajectories were simulated for $t=5000$ with two differential initial conditions corresponding to a small perturbation from the two symmetric equilibrium states: $(x(0),X(0),Y(0),Z(0))=(0,\pm\sqrt{R},\pm\sqrt{R},0)+\xi$ where the same random perturbation $\xi$ in the range $[-10^{-3},10^{-3}]$ was used for both initial conditions. (b) Maximal Lyapunov exponent (MLE) of the simulated trajectories as a function of $R$ where trajectories for which $\text{MLE}>0.01$ were classified as chaotic (purple) in panel (a). (c) The different colors represent the distinct types of particle trajectories. Red denotes self-trapped oscillating WPE that corresponds to a symmetric limit cycle in the phase space, yellow denotes runaway oscillating WPE corresponding to an asymmetric limit cycle in the phase space and purple denotes irregular chaotic motion of the WPE arising from dynamics on a strange attractor in the phase space.}
\label{Fig: dynamics}
\end{figure*}

Furthermore, to explore the very high memory regime, we consider the limit where the individual waves generated by the particle don't decay in time, i.e.~the \emph{infinite memory} limit where $\tau \rightarrow \infty$. This situation is schematically depicted in Fig.~\ref{Fig: schematic}(a). In this limit, the system of equations in Eq.~\eqref{lorenz} further simplify to
\begin{equation}
    \label{diffless lorenz}
    \begin{split}
    \dot{x}_d&=X, \\
    \dot{X}&=Y-X, \\
    \dot{Y}&=XZ, \\
    \dot{Z}&=R-XY.
  \end{split}
\end{equation}
 One can identify that the last three ODEs of \eqref{diffless lorenz} constitute a simplified Lorenz system known as the diffusionless Lorenz equations (DLEs) that has been well studied by \citet{difflesslorenz2000}. It is algebraically simpler than the classic Lorenz system~\citep{Lorenz1963} and one of the algebraically simplest dynamical system that exhibits chaos~\citep{sprott2011}. Thus, the dynamics of a $1$D WPE with a sinusoidal wave form in the limit of infinite memory are described by DLE with a single dimensionless wave-amplitude parameter $R$. However, one should note that in addition to $X=\dot{x}_d$, $Y$ and $Z$ also depend on the particle's motion via \eqref{eq: Y Z}. Hence, for the system in \eqref{diffless lorenz}, the initial condition is restricted by
 \begin{align}\label{eq: Y Z init}
    Y(0)&=R \lim_{\tau\xrightarrow{}\infty} \int_{-\infty}^{0} \sin\left( x_d(0) - x_d(s) \right)\,\text{e}^{\frac{s}{{\tau}}}\,\text{d}s,\\ \nonumber
    Z(0)&=R \lim_{\tau\xrightarrow{}\infty} \int_{-\infty}^{0} \cos\left( x_d(0) - x_d(s) \right)\,\text{e}^{\frac{s}{{\tau}}}\,\text{d}s.
\end{align}
One can also have another viewpoint for the system in \eqref{diffless lorenz} where the variables of the DLE system can be thought of as independent from the particle's dynamical variables and the restriction on initial conditions can be relaxed. The first equation in \eqref{diffless lorenz} can then be interpreted as an equation of motion that connects the particle dynamics with an independent DLE system i.e. an overdamped active particle driven by the $X$ variable of the DLE system. The DLE system may be thought of as a representation of the active particle's internal complexity and DLE variables $(X,Y,Z)$ form the internal state-space of the particle (see Fig.~\ref{Fig: schematic}(b)). From this viewpoint, our system forms an example of an attractor-driven particle~\citep{Valaniattractormatter2023}, i.e.~an overdamped active particle driven by internal low-dimensional DLE system. A practical implementation of this attractor-driven particle could be an autonomous mobile robot that solves the DLE using an onboard CPU or a chaotic circuit and controls the motion of its wheels based on the output of its internal state $(X,Y,Z)$~\citep{arxiv.2306.06609,5430468,976022,Robot1,Robot2,Robot3,Robot4}. Although the focus of the results presented in this paper will mainly be from the viewpoint of WPE, we take the liberty to choose $Y(0)$ and $Z(0)$ independently of the particle's history to explore fully the possible range of behaviors exhibited by the dynamical system. Hence, some of these initial conditions may not be realized for WPE but will be relevant for the attractor-driven particle. For the numerical simulations presented in this paper, the system of ODEs in Eq.~\eqref{lorenz} is solved in MATLAB using the inbuilt solver ode45. %\textbf{While doing simulations I will need to be careful whether my solutions have converged!! Especially for chaotic vs high periodic cases!!! I can verify this by soving using different tol for ode45 and/or trying RK4 with a fixed time step or other stiff methods or implicit methods if the parameter are too small.} %We note that the initial conditions for these simulations correspond to fixed points of the dynamical system; however, there is an intrinsic perturbation provided by numerical round-off errors in MATLAB resulting in unsteady dynamics when the fixed points are unstable. Moreover, for all results presented in this paper, the system was simulated till $t=2000$.

\section{Equilibrium states and their linear stability}\label{sec: lin stab}

At finite memory, there are two equilibrium states of the WPE, a stationary state and a steady walking state~\citep{ValaniUnsteady}. The system in \eqref{diffless lorenz} has no equilibrium points and hence no stationary states for $R>0$. However, excluding the first equation and solving for equilibrium states in \eqref{diffless lorenz} one gets the following steady walking equilibrium state:
\begin{equation}\label{eq: steady eq}
(X,Y,Z)=(\pm\sqrt{R},\pm\sqrt{R},0).
\end{equation}
Note that for this steady walking state, the dimensionless wave height $Z$ vanishes at the particle location but the wave gradient and hence the wave-memory force $Y$ is non-zero resulting in steady walking. In fact, comparing this solution to the finite memory steady walking solution~(see Appendix~\ref{sec: equilibrium})
$$(X,Y,Z)=\left(\pm\sqrt{R-\frac{1}{\tau^2}},\pm\sqrt{R-\frac{1}{\tau^2}},\frac{1}{\tau}\right),$$
we find that the infinite memory limit corresponds to a maximum wave-memory force and a maximum steady walking speed that the WPE can achieve for a given $R$.

To deduce the linear stability of the steady WPE with infinite wave-memory, we apply a small perturbation to this equilibrium state $(X,Y,Z)=(\pm\sqrt{R},\pm\sqrt{R},0)+\epsilon(X_1,Y_1,Z_1)$, where $\epsilon>0$ is a small perturbation parameter. This results in the following linear system that governs the leading order evolution of perturbations:
\begin{gather*}
 \begin{bmatrix} 
 \dot{X}_1 \\
 \dot{Y}_1 \\
 \dot{Z}_1 
 \end{bmatrix}
 =
  \begin{bmatrix}
-1 & 1 & 0 \\
0 & 0 & \pm\sqrt{R} \\
\mp\sqrt{R} & \mp\sqrt{R} & 0
 \end{bmatrix}
  \begin{bmatrix}
  {X}_1 \\
 {Y}_1 \\
 {Z}_1 
 \end{bmatrix}.
\end{gather*}
The linear stability is determined by the eigenvalues of the right-hand-side matrix. This results in the following characteristic polynomial equation to be solved for the eigenvalues $\lambda$ which determine the growth rate of perturbations:
$$\lambda^3 + \lambda^2 + R\lambda + 2R =0.$$
By using Descartes' rule of sign, we either have (i) one negative real eigenvalue and a complex conjugate pair or (ii) three negative real eigenvalues. We can get further clarity by finding the discriminant of this cubic eigenvalue equation which gives 
$$\Delta=-R (4 R^2 + 71 R + 8).$$
Since, this is always negative for $R>0$, we always have a complex conjugate pair of eigenvalues and thus we have one negative real eigenvalue and a complex conjugate pair whose real part is positive~\citep{difflesslorenz2000}~(see Appendix~\ref{sec: equilibrium}).  %\textit{Here can I say something about the real part of the complex conjugate pair? That it is always positive?}

Thus the pair of equilibrium points in Eq.~\eqref{eq: steady eq} are unstable, specifically a pair of saddle-focus~\citep{jackson_1990}, implying that the steady motion of WPE at infinite memory is always unstable to small perturbations. Moreover, the presence of a complex conjugate pair of eigenvalues hints at oscillatory instability of the steady WPE. From the viewpoint of an attractor-driven particle, the instability of the steady state implies that the internal state $(X,Y,Z)$ never settles onto a fixed point and is always changing with time. We now proceed to numerically explore the various unsteady dynamical behaviors arising in the parameter-space of the system.

\section{\label{sec: PS space} Dynamics in the parameter space}

We start by presenting the main types of unsteady dynamical behaviors observed in the motion of the WPE as a function of the dimensionless wave-amplitude parameter $R$. At first it may appear that in the infinite-memory limit, the non-decaying nature of the waves can potentially lead to an ever increasing amplitude of the underlying wave-field via constructive interference of individual waves, i.e.~$|Z|\xrightarrow{} \infty$, and consequently an ever increasing magnitude of wave-memory force, $|Y|\xrightarrow{} \infty$, and particle speed, $|X|\xrightarrow{} \infty$. However, the divergence of the dynamical flow, $\nabla \cdot (\dot{X},\dot{Y},\dot{Z})=-1$, is negative, hence phase-space volume elements contract with time resulting in the $(X,Y,Z)$ phase-space trajectories settling onto an attractor of the system and remaining bounded. Since the steady states of the system are always unstable, we only find unsteady behaviors of the WPE that result in limit cycle attractors and strange attractors in phase-space.

Figure~\ref{Fig: dynamics}(a) shows a bifurcation diagram as a function of the parameter $R$ which plots the wave-memory force on the particle, $Y_n=Y(t_n)$, sampled at times $t_n$ that correspond to the particle's instantaneous velocity being zero i.e. $X(t_n)=0$. The corresponding maximal Lyapunov exponent (MLE) as a function of $R$ is shown in Fig.~\ref{Fig: dynamics}(b). A positive MLE hints at the presence of chaos. The bifurcation diagram is colored based on the following three distinct types of unsteady WPE dynamics: (i) self-trapped oscillating WPE (red) where the WPE undergoes back-and-forth oscillations about a fixed point, (ii) runaway oscillating WPE (yellow) where the WPE undergoes inline oscillations along with a net drift, and (iii) irregular WPE (purple) that exhibit chaotic walks. We can associate these motions of WPE to the corresponding dynamics and geometry in the $(X,Y,Z)$ phase-space of DLE~(see Fig.~\ref{Fig: dynamics}(c)). The self-trapped oscillating WPE corresponds to a limit cycle in phase-space with a symmetric geometry while the runaway oscillating WPE corresponds to a limit cycle in phase-space with asymmetric geometry. Runaway oscillating WPE occur in symmetric pairs that correspond to the net drift of the WPE in the positive or negative direction. Irregular chaotic WPE correspond to dynamics on a strange attractor in phase space. We now proceed to explore the dynamics as a function of $R$ in detail by dividing the parameter-space into three regions: (i) small wave-amplitude regime $(R<0.2)$, (ii) intermediate wave-amplitude regime $(0.2<R<2)$ and (iii) large wave-amplitude regime $(R>2)$.

%\begin{figure}[h]
%\centering
%\includegraphics[width=\columnwidth]{Bifurcations_yzero_inf_mem_walker_R_0.02_to_0.20_dR_0.00_tend_2000.00_NEW.png}
%\caption{Bifurcation diagram from $R=0.02$ to $R=0.2$}
%\label{Fig: bif1}
%\end{figure}

%\begin{figure}[h]
%\centering
%\includegraphics[width=\columnwidth]{Bifurcations_yzero_inf_mem_walker_R_0.20_to_1.00_dR_0.00_tend_2000.00_NEW.png}
%\caption{Bifurcation diagram from $R=0.2$ to $R=1$}
%\label{Fig: bif2}
%\end{figure}

%\begin{figure}[h]
%\centering
%\includegraphics[width=\columnwidth]{Bifurcations_yzero_inf_mem_walker_R_1.00_to_10.00_dR_0.01_tend_2000.00_NEW.png}
%\caption{Bifurcation diagram from $R=1$ to $R=10$}
%\label{Fig: bif3}
%\end{figure}

%\begin{figure}[h]
%\centering
%\includegraphics[width=\columnwidth]{Basin_of_attraction-R_0p22.png}
%\caption{Basin of attraction for $R=0.22$}
%\label{Fig: basin1}
%\end{figure}

%\begin{figure}[h]
%\centering
%\includegraphics[width=\columnwidth]{Basin_of_attraction-R_0p10.png}
%\caption{Basin of attraction for $R=0.10$}
%\label{Fig: basin2}
%\end{figure}

\subsection{\label{sec: small R} Small wave-amplitude regime}

\begin{figure*}
\centering
\includegraphics[width=2\columnwidth]{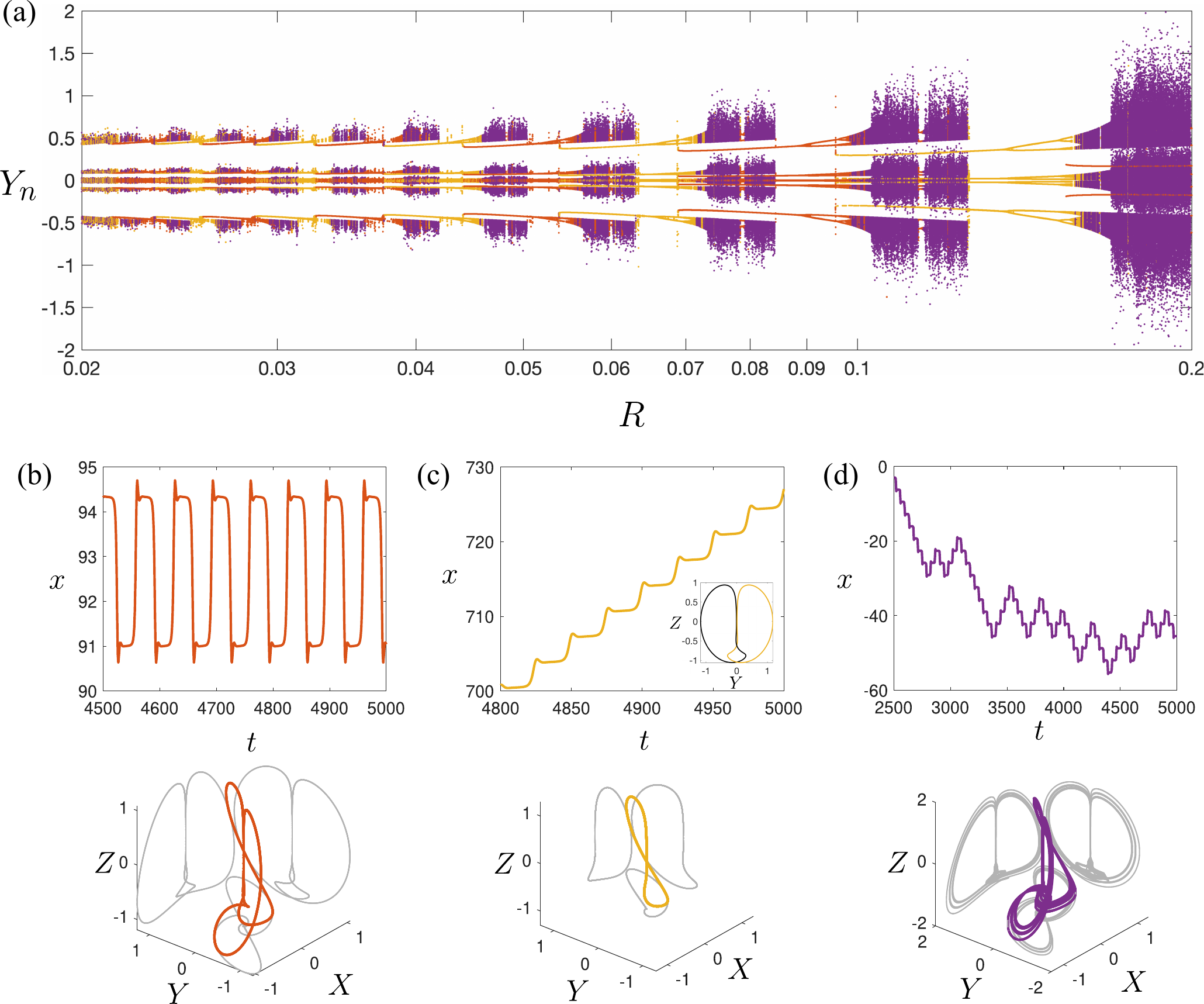}
\caption{Dynamics in the small wave-amplitude regime. (a) Bifurcation diagram showing the wave-memory force $Y_n$ acting on the WPE when its instantaneous velocity is zero i.e. $X=0$. The different colors represent the qualitatively different kind of particle trajectories for the WPE. (b) Sample space-time trajectory (top panel) and phase-space attractor (bottom panel) for self-trapped intermittent WPE (red) at $R=0.08$, (c) runaway intermittent WPE (yellow) at $R=0.1$ and (d) irregular intermittent WPE (purple) at $R=0.082$.}
\label{Fig: intermittent dynamics PS}
\end{figure*}

\begin{figure}
\centering
\includegraphics[width=\columnwidth]{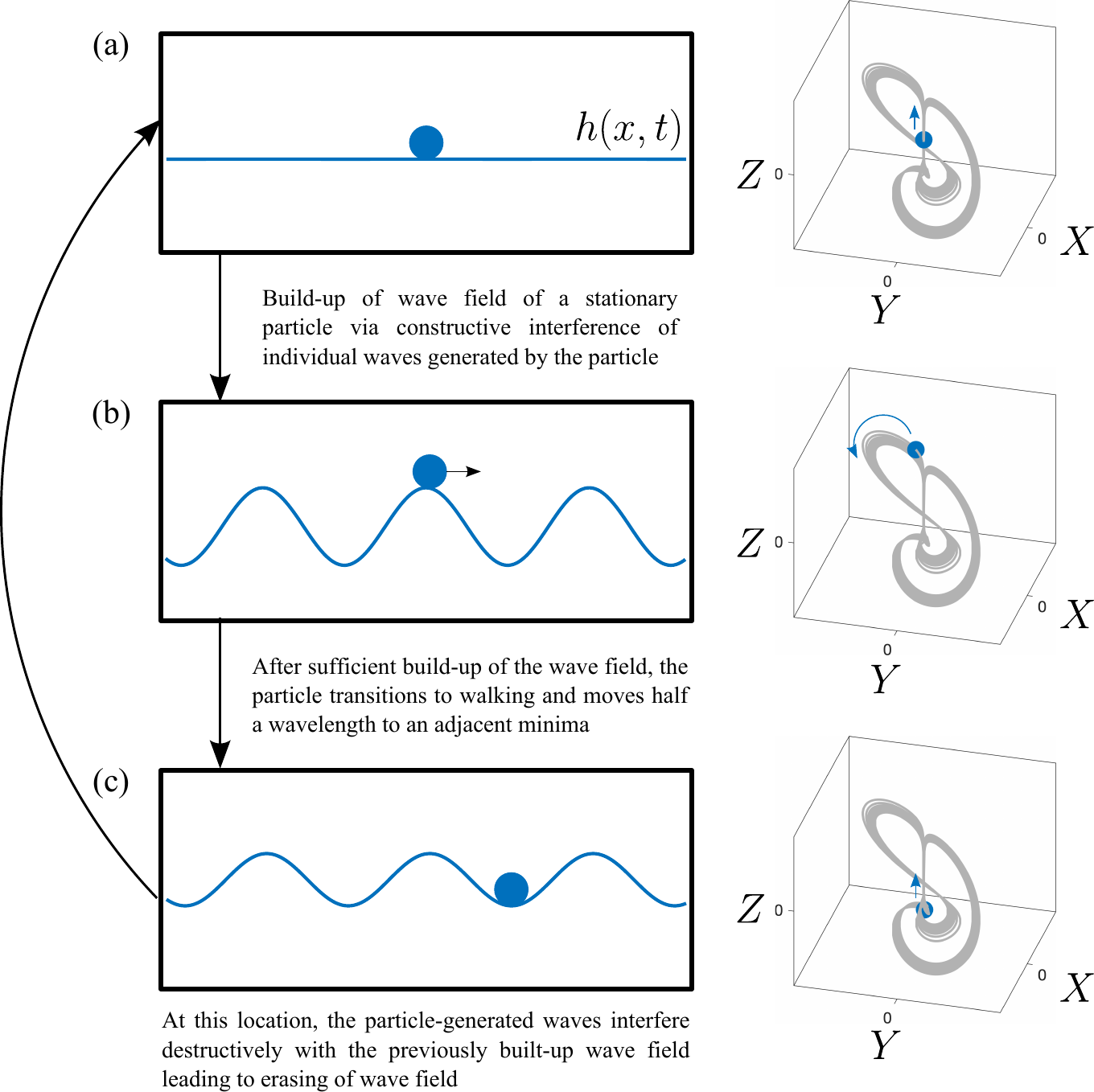}
\caption{(Multimedia view) Physical mechanism of intermittent dynamics of the WPE in the small $R$ regime and its connection to the phase-space dynamics. (a) An initially stationary particle with no wave field slowly builds-up its wave field $h(x,t)$ via constructive interference of individual small amplitude cosine waves centered at the location of the particle. This corresponds to the phase-space trajectory slowly climbing along the $Z$ axis with $Z>0$. (b) After sufficient build-up of the wave field, the stationary particle transitions to walking and swiftly moves to an adjacent minima of its self-generated wave field. This corresponds to the phase-space trajectory quickly traversing one of the ``wings" (left or right) of the phase-space attractor. (c) At this minima, the particle generated cosine waves interfere destructively with the previously built-up wave field, leading to erasing of the wave field. This corresponds to the phase-space trajectory again climbing the $Z$ axis with $Z<0$. This cycle repeats and results in intermittent motion of the WPE.}
\label{Fig: intermittent dynamics mechanism PS}
\end{figure}

We start by exploring in detail the regime of small wave-amplitude parameter $R$. Figure~\ref{Fig: intermittent dynamics PS}(a) shows a detailed bifurcation diagram, similar to the one in Fig.~\ref{Fig: dynamics}(a), but focused on the region $0.02<R<0.2$. We find that this regime is dominated by intermittent dynamics where the trajectory of the WPE alternates between long stationary phases and short walking phases. Three different types of motion described in the previous section are realized in this regime with intermittent dynamics: (i) self-trapped intermittent WPE as shown in Fig.~\ref{Fig: intermittent dynamics PS}(b), (ii) runaway intermittent WPE as shown in Fig.~\ref{Fig: intermittent dynamics PS}(c), and (iii) irregular intermittent WPE as shown in Fig.~\ref{Fig: intermittent dynamics PS}(d). As shown in the inset of Fig.~\ref{Fig: intermittent dynamics PS}(b), two symmetric pair of solutions exist for runaway intermittent WPE that correspond to the net drift of the particle in the positive (yellow) and negative (black) direction. We see from the crossing of the two asymmetric limit cycles that they are topologically linked in phase-space. 

From the bifurcation diagram in Fig.~\ref{Fig: intermittent dynamics PS}(a), we can see that the three types of intermittent motion are intricately interwoven in what appears to be a self-similar period-doubling bifurcation structure with an increasing number of bifurcations squeezed into an infinitely thin region as $R\xrightarrow{} 0$. We refer the interested reader to the work of \citet{difflesslorenz2000} who derived analytical approximations of such self-similar bifurcations arising in this regime. We further observe in Fig.~~\ref{Fig: intermittent dynamics PS}(a) that the periodic attractors for these self-similar bifurcations alternate between symmetric (red) and asymmetric (yellow) limit cycles. Self-similar bifurcations have also been recently reported in certain regimes of the classic Lorenz system~\citep{CHEN2023113651}; this system is connected to $1$D WPE dynamics at finite memory~\citep{ValaniUnsteady,Valanilorenz2022}. The bifurcation diagram also reveals multistability in this system as evident from the presence of multiple attractors (multiple colors) at the same $R$ value. We will discuss this aspect of multistability and the corresponding basin of attraction in some detail in Sec.~\ref{sec: Multistab}.

\begin{figure*}
\centering
\includegraphics[width=2\columnwidth]{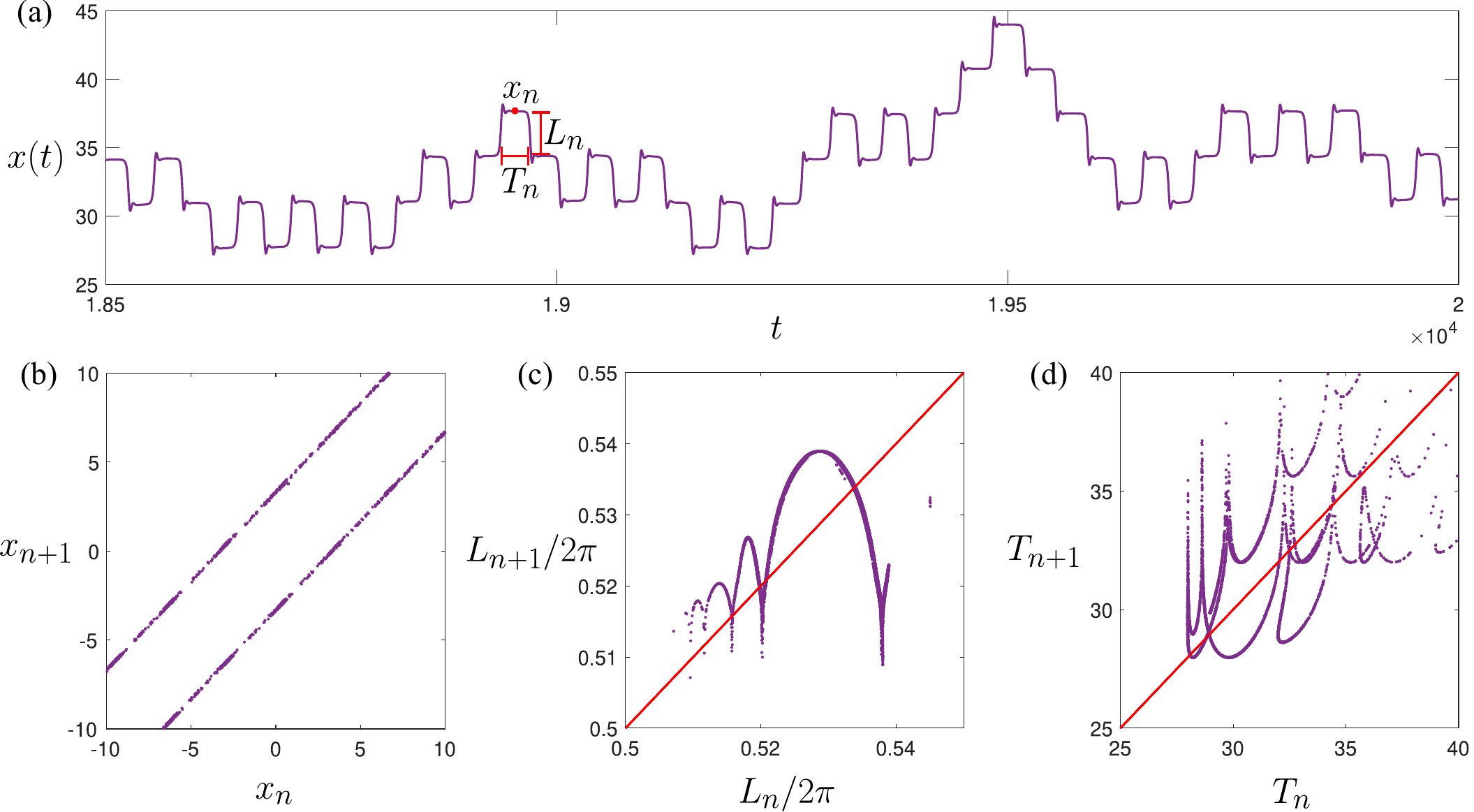}
\caption{Chaotic aspects of irregular intermittent WPE dynamics for $R=0.11$. (a) Space-time trajectory of the particle. (b) One-dimensional return map of the position of the particle when it is in the stationary phase at the $n$th step, $x_n$, versus its position in the stationary phase at the ($n+1$)th step, $x_{n+1}$. (c) One-dimensional return map of the length of the $n$th step, $L_n$, versus the length of the ($n+1$)th step, $L_{n+1}$, both scaled by the wavelength $2\pi$ of the particle-generated waves. (d) One-dimensional return map of the time spent in the stationary state at the $n$th step, $T_n$, versus at the ($n+1$)th, $T_{n+1}$. The red lines in (c) and (d) are diagonal lines whose intersection with the purple curves corresponds to, respectively, $L_{n+1}=L_{n}$ and $T_{n+1}=T_{n}$ and they determine equilibria of the map dynamics.}
\label{Fig: intermittent chaotic map}
\end{figure*}

We proceed by providing a physical mechanism for the intermittent motion of the WPE realized in this small $R$ regime and connect it to the phase-space dynamics of the DLE system. Recall that $R$ is a dimensionless wave-amplitude parameter and in the small $R$ regime, it takes a long time for the particle to build up its overall wave field via superposition of small amplitude waves. To understand this further, we consider our dynamical system for the state 
$$X=Y=0,$$ and $$\dot{Z}=R,$$ or alternatively $Z=Rt$. This corresponds to a stationary WPE with zero velocity ($X=0$) and zero horizontal wave-memory force ($Y=0$), but the wave height beneath the particle is increasing at a constant rate $R$ starting from a zero wave height $Z=0$. The corresponding Jacobian of the dynamical system for this state results in the following approximate eigenvalues for $Rt\ll 1$:
\begin{equation*}
    \lambda \approx 0,-1-Rt,Rt
\end{equation*}
with the corresponding eigenvectors $(0,0,1), (-1/(Rt),1,0)$ and $(1/(1+Rt),1,0)$, respectively. Thus, the small positive eigenvalue of $Rt$ has an eigenvector approximately $(1,1,0)$ corresponding to destabilization of the stationary state at a rate proportional to $R$. Hence, even though the stationary state of the particle is unstable for $R>0$, for $0<R\ll 1$ it takes a long time for the particle to build up its wave field and destabilize from stationary state contributing to long stationary phases in the intermittent motion. Figure~\ref{Fig: intermittent dynamics mechanism PS} explains this physical mechanism of intermittent motion and relates it to the corresponding phase-space dynamics. Initially, the WPE is stationary with no wave field and it slowly builds its wave field through constructive interference of small amplitude sinusoidal waves that the WPE generates at each instant (see  Fig.~\ref{Fig: intermittent dynamics mechanism PS}(a)). This corresponds to the slow motion away from the dynamical state $(X,Y,Z)=(0,0,0)$ along the $Z$-axis in the direction $Z>0$ at a rate proportional to $R$ in phase-space. Once the wave field builds up sufficiently, the WPE starts moving (left or right) and traps itself in an adjacent trough of its self-generated wave field (see Fig.~\ref{Fig: intermittent dynamics mechanism PS}(b)). This corresponds to the phase-space trajectory quickly traversing left or right ``wing" of the Lorenz-like attractor in phase-space. The trapped WPE continues to generate new waves at the location of this trough, resulting in destructive interference and the overall wave field amplitude starts decreasing i.e. the particle erases its self-generated wave field (see Fig.~\ref{Fig: intermittent dynamics mechanism PS}(c)). This corresponds to the phase-space trajectory again slowly climbing along the $Z$ axis with $Z<0$. After the WPE erases its wave field, it starts building a new wave field at this location until it starts moving again and repeats this cycle. Thus, this process results in intermittent motion of the WPE with two distinct phases: (i) a \emph{slow} stationary phase where the WPE is erasing and building its wave field, and (ii) a \emph{fast} walking phase where the WPE moves and takes a step of nearly half the wavelength i.e. from the peak of its wave field to a nearby trough. The half-wavelength step of the intermittent WPE is a reflection of an orbit around one ``wing" of the Lorenz-like attractor in the phase-space of the dynamical system. This also shows that the infinite-memory of the particle-generated waves is a red herring in this regime; since the WPE periodically erases its wave-memory during intermittent dynamics, the waves generated in the distant past can have little effect on the particle motion at present.

We have observed similarities between the intermittent dynamics described here and the {stop-and-go motion} of superwalking droplets~\citep{superwalker,ValaniSGM}. Superwalkers~\citep{superwalker, superwalkernumerical} are bigger and faster walking droplets that emerge when the bath is vibrated at two frequencies simultaneously, namely a particular frequency and half of that frequency, along with a constant phase difference. By detuning the two driving frequencies by a small amount, one can get the phase difference to drift slowly in time. This detuned two-frequency driving results in a novel walking motion for superwalkers known as stop-and-go motion~\citep{superwalker,ValaniSGM}. The stop-and-go motion of droplets, enabled by the varying phase difference, results in periodic traversals of the stationary and walking regimes in the parameter-space of the physical system. In their simulations of stop-and-go motion, \citet{ValaniSGM} reported three different types: (i) back-and-forth, (ii) forth-and-forth and (iii) irregular. The particle trajectories of these stop-and-go motions and the three intermittent motions observed in our system (self-trapped intermittent, runaway intermittent and irregular intermittent) are very similar despite being different systems. The stop-and-go motion of superwalkers is a complex nonlinear phenomenon with multiple time scales coming into play such as the bouncing time scale of the droplet, the memory time scale associated with decay of droplet-generated waves, the even longer time scale introduced by the detuning and the time scale of the inertial response of the droplet. Conversely, the mechanism for these intermittent motion for our WPE does not required any external parametric driving of the system between stationary and walking states, as done in stop-and-go motion. The intermittent motion of our WPE at infinite memory is an emergent phenomena arising from a combination of a slow instability of the stationary state and the WPE trapping itself in a nearby trough when walking.

We note that \citet{Durey2020} also reported intermittent motion of the WPE (called jittering modes in their paper) in their numerical simulations of the integro-differential equation of motion with a Bessel wave form in the high-memory regime. They rationalized the mechanics of growth-relaxation process of intermittent motion in terms of a linearized integro-differential equation model during the growth stage of the wave field and an overdamped particle moving in a static potential during the relaxation stage. The similarities in intermittent dynamics suggest that our simple WPE with sinusoidal wave form and infinite memory can successfully capture at least qualitatively these intricate states observed with a more realistic Bessel wave form. Moreover, our simple ODE model sheds light on the dynamical processes in $3$D phase-space that result in intermittent trajectories of the WPE in physical space-time. It further suggests that a low-dimensional attractor might be governing the dynamics of the WPE with Bessel wave form in this regime. \citet{Durey2020} further analyzed the irregular intermittent WPE motion by modeling it as a stochastic, discrete-time, Markovian jump process, where the particle can move left or right with equal probability and a fluctuating step-length. They showed that this model showed consistent results with their simulations with multimodal statistics at intermediate timescale and Gaussian distribution at long times in the particle's position distribution. In the remainder of this section, we explore chaotic and statistical properties of irregular intermittent WPEs using our ODE model and compare some of these results with that of \citet{Durey2020} obtained for the Bessel function wave form.

We start by exploring the chaotic aspects of irregular intermittent WPE. A typical trajectory is shown in Fig.~\ref{Fig: intermittent chaotic map}(a) for $R=0.11$. To understand the chaotic behavior, we plot $1$D return maps of different quantities associated with the trajectory of irregular intermittent WPE. A $1$D return map that plots the location of particle $x_n$ in the $n$th stationary phase versus the ($n+1$)th stationary phase is shown in Fig.~\ref{Fig: intermittent chaotic map}(b). One observes two parallel line structure in this map indicating that the map is multi-valued. This is consistent with the intermittent irregular trajectory since at a given location in the stationary phase, $x_n$, the particle can unpredictably either take the next step, $x_{n+1}$, to its left or to its right. We note that the structure of this return map and the intermittent irregular WPE trajectories are reminiscent of pseudolaminar chaotic diffusion~\citep{pseudolaminar2023} where a time-series with constant-value laminar phases is periodically interrupted by chaotic bursts. This is different from laminar chaos arising in time-delay systems with periodically varying delay where also a similar time series is encountered but in addition the levels of laminar phases in the time series are related by a simple and robust one-dimensional map~\citep{Muller2018,Albers2022}. Next, we analyze the return map of the step length $L_n$ (scaled by the wavelength $2\pi$) of consecutive steps resulting in Fig.~\ref{Fig: intermittent chaotic map}(c). We make two observations: (i) all the steps are nearly constant and slightly bigger than half the wavelength, and (ii) the variations in the step length are well captured by this $1$D return-map with a well-defined structure. Hence, we see evidence of low-dimensional chaos in the variations in step-length during intermittent irregular trajectories. However, instead of length of the step, if one plots the return map of consecutive durations $T_n$ of time spent in the stationary phase then one gets a more complicated multi-valued map as shown in Fig.~\ref{Fig: intermittent chaotic map}(d). It would be interesting to compute these maps for more complete WPE models, such as the Bessel wave form model of \citet{Durey2020} to see if they also show signatures of low-dimensional chaos for these quantities.

\begin{figure*}
\centering
\includegraphics[width=2\columnwidth]{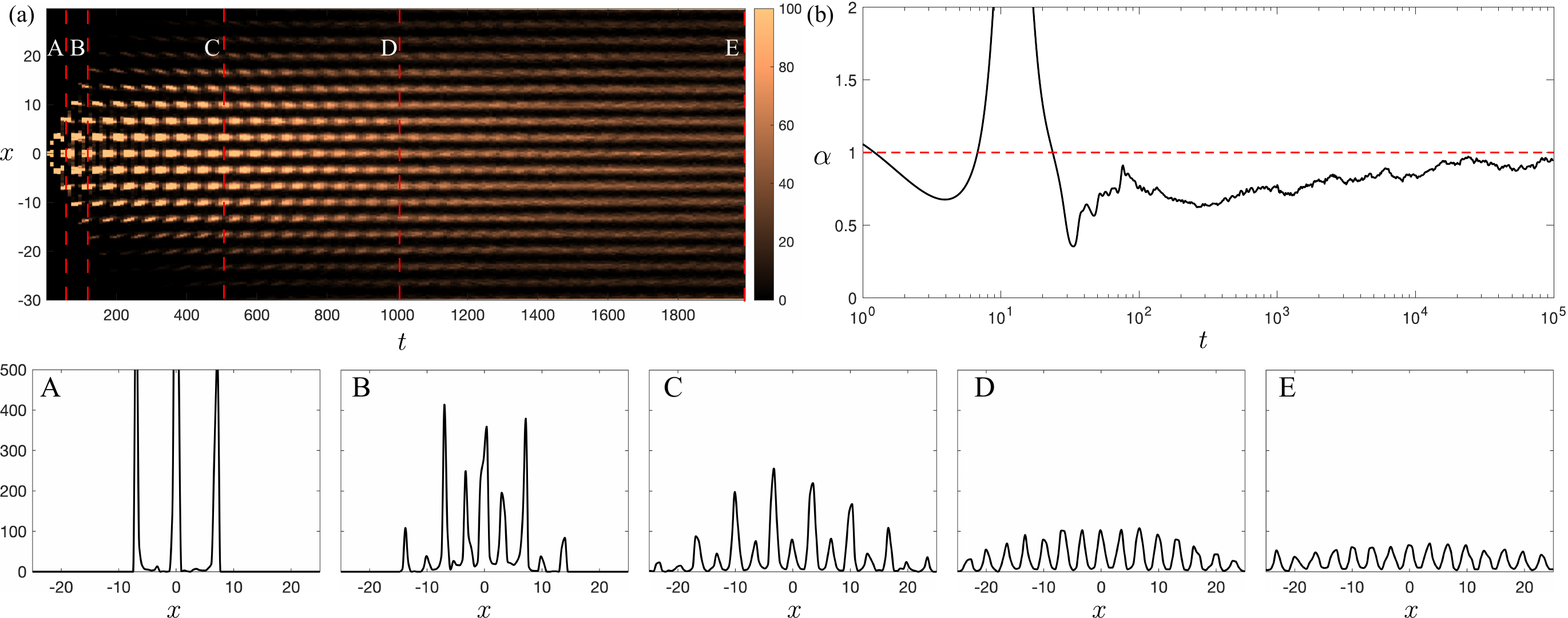}
\caption{Statistical aspects of irregular intermittent WPE dynamics for $R=0.11$. (a) Position distribution for irregular intermittent WPE as a function of time. The distribution was calculated by sampling $3000$ intermittent irregular trajectories starting with initial conditions $(x_d(0),X(0),Y(0),Z(0))=(0,X_0,0,0)$ where $X_0$ was uniformly sampled randomly from the interval $[-0.5,0.5]$. Probability distributions are shown (red dashed lines) at (A) $t=50$, (B) $t=100$, (C) $t=500$, (D) $t=1000$ and (E) $t=2000$. (b) Time-dependent diffusion exponent $\alpha$ as a function of time, calculated from an ensemble of $750$ trajectories of intermittent irregular WPEs at $R=0.11$ simulated for $t=10^5$.}
\label{Fig: Probability distribution}
\end{figure*}

\begin{figure*}
\centering
\includegraphics[width=2\columnwidth]{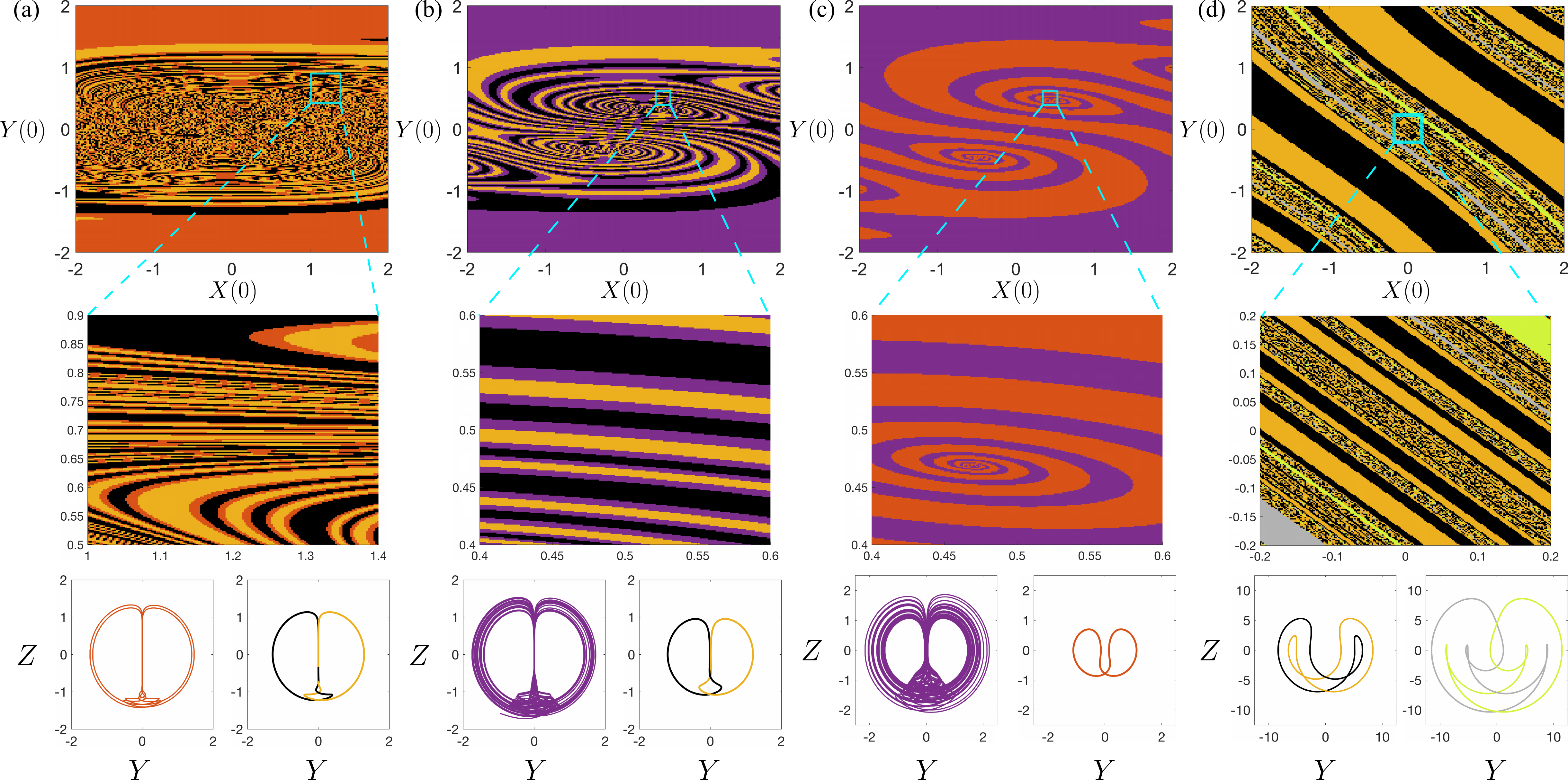}
\caption{Multistability and basin of attraction. Basin of attraction in the initial condition space $(X(0),Y(0))$ (top and middle panels) and the corresponding attractors in phase-space (bottom panel) for the coexistence of (a) self-trapped oscillating and runaway oscillating WPEs at $R=0.06$, (b) irregular and runaway oscillating WPEs at $R=0.11$, (c) irregular and runaway oscillating WPEs at $R=0.22$ and (d) 
 two different types of runaway oscillating WPEs at $R=6$. The initial values of $x_d$ and $Z$ were fixed to zero.}
\label{Fig: multi}
\end{figure*}

We now explore the statistical features of these nearly constant step-length chaotic walks. We have calculated the position distribution of the particle in this irregular intermittent regime as a function of time as shown in Fig.~\ref{Fig: Probability distribution}(a). This was done by initiating $3000$ WPEs at $x_d(0)=0$ with a uniform velocity ($X$) distribution in the range $[-0.5,0.5]$. We note that only those trajectories that displayed intermittent irregular behavior were included in the probability distribution since the system exhibits multistability~(see Fig.~\ref{Fig: multi}(a)). From Fig.~\ref{Fig: Probability distribution}(a) we observe a wave-like probability distribution that persists for a long time. The long persistence of the wave-like distribution can be attributed to the coherence of the sinusoidal waves. Since each step taken by the WPE is nearly of half the wavelength i.e. $\pi$, with a narrow distribution in uncertainty~(see Fig.~\ref{Fig: intermittent chaotic map}(c)), it may take a very long time for these small differences in step-length uncertainties to accumulate and for the sharp peaks to diffuse. However, as it can be seen in bottom panels (A-E) of Fig.~\ref{Fig: Probability distribution}, the envelope of the distribution diffuses with time and the sharp wave-like features decay. Thus, we observe that wave-like features in the probability distribution persist for long time but the spreading of the overall envelope due to the diffusive nature of trajectories results in these wave-like features diminishing with time. For the WPE dynamics in this regime with a Bessel wave form studied by \citet{Durey2020}, they observed that wave-like features diffused relatively quickly into a Gaussian-like distribution which then spreads over space. The relatively early suppression of wave-like features in the probability distribution with Bessel wave form might be due to phase shifts between consecutive peaks of the Bessel function (as compared to sinusoidal waves) in combination with spatial decay. Such features may result in larger fluctuations in the step-length of intermittent irregular WPEs and hence the distribution transitions early from multimodal to Gaussian compare to our sinusoidal wave form. We also observe another feature from Fig.~\ref{Fig: Probability distribution}(a) that this probability distribution oscillates at small and intermediate time-scales i.e. at a fixed location $x$, the probability distribution is oscillating with time. This is due to the discrete nature of the intermittent trajectories where a given location in space is occupied and unoccupied by different intermittent irregular walkers of nearly constant step-length. 

The diffusive behavior of intermittent irregular WPEs can be characterized by calculating how the mean squared displacement (MSD) scales with time, i.e., $\text{MSD}=\langle(x_d(t)-x_d(0))^2\rangle \sim t^\alpha$ with $\alpha$ being the diffusion exponent. To quantify this, we define a time dependent diffusion exponent $\alpha(t)=\text{d}(\text{log}(\text{MSD}))/\text{d}(\text{log}(t))$ and plot it as a function of time as shown in Fig.~\ref{Fig: Probability distribution}(b). We observe subdiffusion ($0<\alpha<1$) for intermediate time scales and the WPE appears to be approaching normal diffusion i.e. $\alpha\xrightarrow{}1$ asymptotically. This was also observed by \citet{Durey2020} for intermiitent irregular WPE dynamics with Bessel wave form.

%Figure~\ref{Fig: Probability distribution}(b) shows the distrubition of velocity. We find that at long times, this relaxes to a central peak at $X=0$ and two small peaks near $X\approx\pm1$. The large peak near $X=0$ signifies that the particles spends most of its time in the stationary state while the two small peaks for nonzero $X$ signify the swift walking between stationary states. Moreover, the oscillations in the PDF also persist here at small and intermediate time scales due to the discrete nature of the trajectories.

From the view-point of an attractor-driven particle, appearance of such wave-like statistical features that persist for long is not common in traditional $1$D active particles. A commonly used minimal model for $1$D active particle is a run-and-tumble particle (RTP). It is an overdamped particle that moves with a constant self-propulsion speed and flips its direction of motion following a constant-rate Poisson process~\citep{PhysRevE.99.032132}. Sometimes Gaussian white noise is also added as an additional stochastic force~\citep{Malakar_2018}. Such RTPs show bimodal position distribution at very short time-scale due to their ballistic motion while at long time scales the distribution approaches a Gaussian. Due to the intermittent nature of the trajectory of our attractor-driven particle, we obtain persistent spatial oscillations and also temporal oscillations at short and intermediate times. Thus, the rich dynamical and statistical features arising from our attractor-driven particle can motivate modeling of new classes of active particles.

\subsection{\label{sec: intermediate R} Intermediate wave-amplitude regime}

\begin{figure}
\centering
\includegraphics[width=\columnwidth]{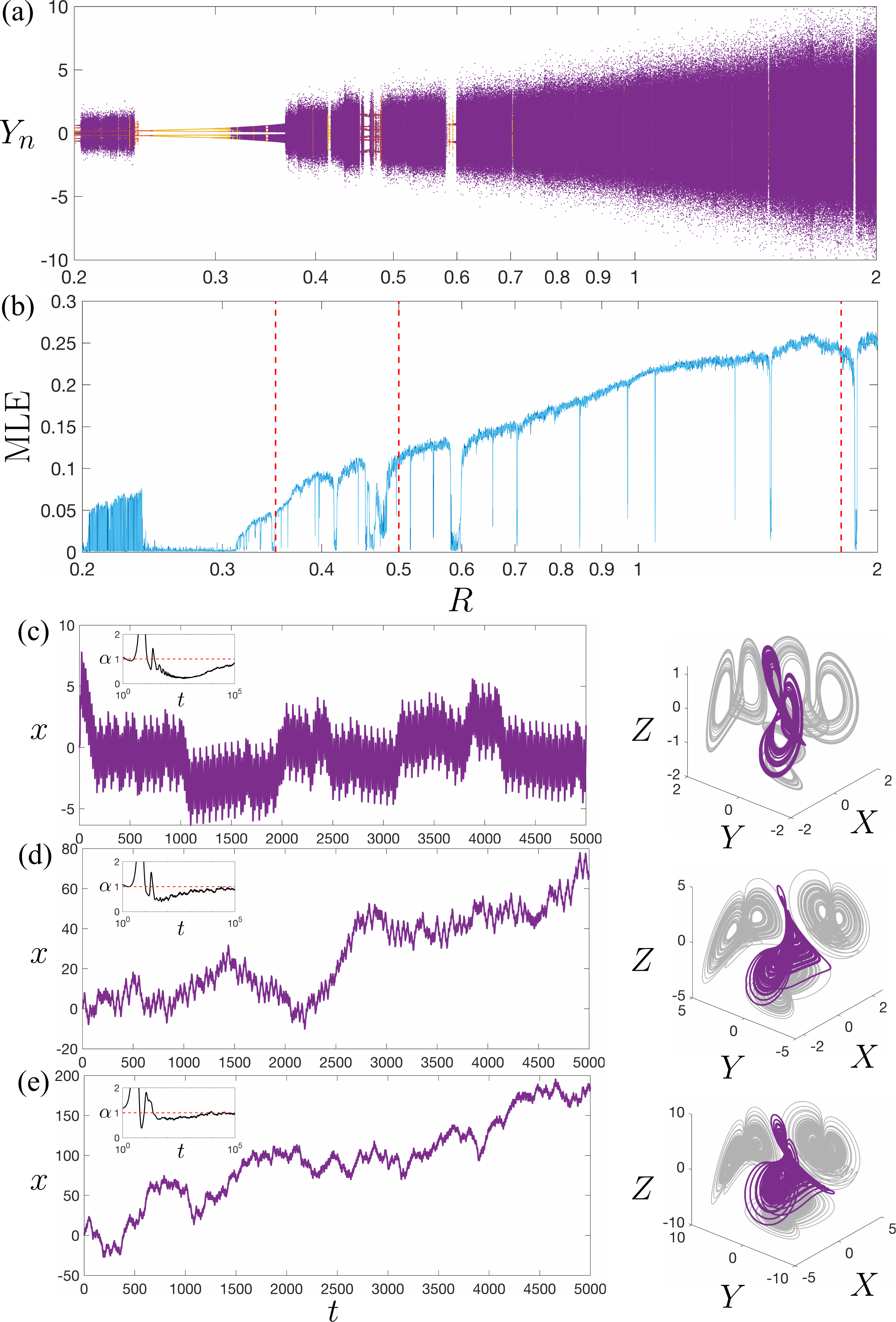}
\caption{Dynamics in the intermediate wave-amplitude regime. (a) Bifurcation diagram in the intermediate $R$ regime ($0.2<R<2$) showing the wave-memory force $Y_n$ acting on the WPE when its instantaneous velocity is zero i.e. $X=0$. The different colors represent the qualitatively different kind of trajectories for the WPE. Red denotes self-trapped oscillating WPE that corresponds to a symmetric limit cycle in the phase space, yellow denotes runaway oscillating WPE corresponding to an asymmetric limit cycle and purple denotes irregular motion of the WPE arising from a strange attractor in the phase space. (b) Maximal Lyapunov exponent (MLE) as a function of $R$ with the red-dashed lines corresponding to panels (c)-(e) which show space-time trajectories (left panel) and phase-space attractor (right panel) for $R=0.35, 0.5$ and $1.8$ respectively. The insets in (c)-(e) shown time-dependent diffusion exponent $\alpha$ as a function of time, calculated from an ensemble of $400$ trajectories simulated for $t=10^5$.}
\label{Fig: intermediate R}
\end{figure}

We now turn to explore the intermediate wave-amplitude regime $0.2<R<2$. This is also the regime of $R$ parameter that would typically correspond to experiments with walkers and superwalkers~\citep{Molacek2013DropsTheory,superwalker} if one can achieve this regime of very high memory and confine the droplet motion to one-dimension e.g. by restricting its motion to a thin annular region~\citep{Filoux2015StringsWaves,PhysRevFluids.5.083601,Rahman2018}. However, we don't expect the dynamics observed here to quantitatively match experiments since we are using an idealized model, but qualitative similarities in trajectories may be realized.

A detailed bifurcation diagram of this regime is shown in Fig.~\ref{Fig: intermediate R}(a) and the corresponding MLE are shown in Fig.~\ref{Fig: intermediate R}(b). We find that this regime mainly comprises of irregular WPEs with small regions of self-trapped WPEs and runaway WPEs. A multi-stable region is observed near $R\approx0.2$ with coexistence of irregular and self-trapped WPEs followed by a region of runaway WPEs near $R\approx0.3$. These runaway WPEs bifurcate into irregular WPEs near $R\approx0.35$. A trajectory of an irregular WPE just after this transition is shown in Fig.~\ref{Fig: intermediate R} (c). Here, we find that the WPE shows subdiffusive behavior for a large range of intermediate timescales and very slowly seems to be approaching asymptotic diffusion (see inset of Fig.~\ref{Fig: intermediate R} (c)). Further increase in $R$ leads to increasing complexity as well as enlarging of the chaotic attractor in phase-space as depicted in Figs.~\ref{Fig: intermediate R}(c)-(e). The increasing physical extent of the attractor is also reflected in the widening of the envelope in the bifurcation diagram in Fig.~\ref{Fig: intermediate R}(a) with increasing $R$. For these trajectories we also find asymptotic diffusion (see inset of Fig.~\ref{Fig: intermediate R} (d-e)) with the diffusion constant typically larger for larger $R$. The MLE typically also increases with increase in $R$ value in this regime~(see Fig.~\ref{Fig: intermediate R}(b)). 

The increasing complexity of the DLE strange attractor also provides a rich set of statistical features for our attractor-driven particle that can be tuned by the control parameter $R$. By varying $R$, one can induced desired transport properties i.e. trapping from self-trapped WPEs, or ballistic motion from runaway WPEs, or subdiffusion and normal diffusion with diffusion coefficients that can be tuned by varying $R$.

%\begin{itemize}
%\item \textbf{Figures for this section:} (1) I can show chaotic trajectories and corresponding phase-space attractors and I can then comment on how they are similar and different from the intermittent dynamics regime. (2) I can potentially look at the diffusive exponent and constant for the chaotic trajectories in this regime. (3) For the periodic trajectories in this regime e.g. $R=6$ has two different types of runaway oscillating WPEs. I can plot and talk about the Basin of attraction and also link it to knots in the limit cycle and reference appropriate books etc..
%\end{itemize}

\begin{figure*}
\centering
\includegraphics[width=2\columnwidth]{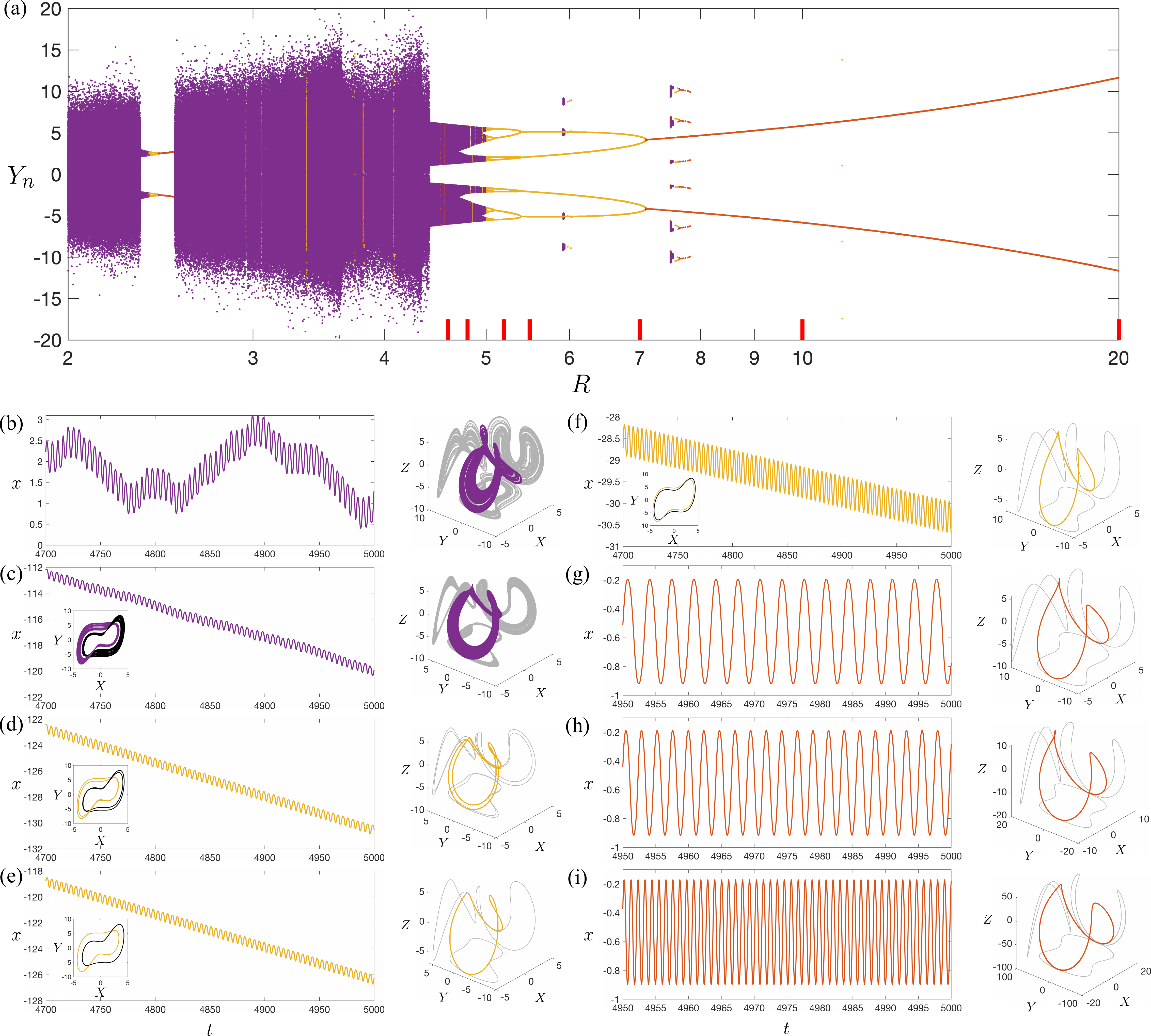}
\caption{Dynamics in the large wave-amplitude regime. (a) Bifurcation diagram in the large $R$ regime ($R>2$) showing the wave-memory force $Y_n$ acting on the WPE when its instantaneous velocity is zero i.e. $X=0$. The red markers at the bottom of this figure correspond to the $R$ values corresponding to panels (b)-(i). The different colors represent the qualitatively different kind of trajectories for the WPE. Red denotes self-trapped oscillating WPE that corresponds to a symmetric limit cycle in the phase space, yellow denotes runaway oscillating WPE corresponding to an asymmetric limit cycle and purple denotes irregular motion of the WPE arising from a strange attractor in the phase space. (b)-(i) Space-time trajectory (left panel) and the phase-space attractor (right panel) for $R=4.6, 4.8, 5.2, 5.5, 7, 10, 20$ and $100$ respectively.}
\label{Fig: high R}
\end{figure*}

\subsection{\label{sec: large R} Large wave-amplitude regime}

We now turn towards the large wave-amplitude regime that corresponds to $R>2$. A detailed bifurcation diagram in this regime is shown for $2<R<20$ in Fig.~\ref{Fig: high R}(a). We see that as $R$ increases in this regime, the irregular WPEs cease near $R=5$. This happens via a period halving bifurcate where irregular WPEs bifurcation into runaway WPEs. These runaway oscillating WPEs further bifurcate into self-trapped oscillating WPEs near $R\approx7$. The self-trapped oscillating WPEs persist as $R\xrightarrow{}\infty$. 

To explore this regime further, we show plots of the space-time trajectories (left) and phase-space attractors (right) for increasing $R$ values in Figs.~\ref{Fig: high R}(b)-(i). At $R=4.6$, a strange attractor exists in phase space with a symmetric structure (Fig.~\ref{Fig: high R}(b)). The symmetry of the attractor implies that the irregular WPE on average has no net displacement. As $R$ increases to $4.8$ (Fig.~\ref{Fig: high R}(c)), a dynamical symmetric breaking~\citep{jackson_1990} takes placing forming a pair of strange attractors (purple and black in the inset of Fig.~\ref{Fig: high R}(c)). The two attractors correspond to a net drift in the positive (black) and negative (purple) direction. This asymmetric strange attractor causes irregular modulations in the oscillations of runaway WPEs. We further observe from the inset of Fig.~\ref{Fig: high R}(c) that the two strange attractors are topologically linked. Further increasing $R$, we obtain a runaway oscillating WPE at $R=5.2$ (Fig.~\ref{Fig: high R}(d)) which undergoes period halving as $R$ increases to $5.5$ (Fig.~\ref{Fig: high R}(e)). For these runaway oscillating walkers, the link between the two attractors is preserved as shown in the insets of Figs.~\ref{Fig: high R}(d) and (e) respectively. On further increasing $R$ to $7$ (Fig.~\ref{Fig: high R}(f)), we see that the phase-space limit cycle of the runaway oscillating WPE becomes less asymmetric corresponding to a smaller drift speed of the WPE. This state eventually transitions to a symmetric limit cycle and one obtains self-trapped oscillating WPE as shown in Fig.~\ref{Fig: high R}(g) for $R=10$. 

From hereon, further increasing $R$ to $20$ (Fig.~\ref{Fig: high R}(h)) and $100$ (Fig.~\ref{Fig: high R}(i)) we find that the extent of the limit cycle in phase space keeps on increasing while the amplitude of oscillations of the WPE in space remain constant and the frequency of oscillations increase. This can be understood as follows (see also Supplemental Video 1): At large $R$, the particle performs self-trapped oscillations between two consecutive peaks of its own wave field. When the particle is near the trough between the two peaks, the particle-generated waves will interfere destructively with the built-up wave field. This will decrease the amplitude of the overall wave field. Conversely, when the particle is near the peaks of the wave field, the particle-generated waves will interfere constructively with the built-up wave field and the overall wave field amplitude will increase. Now, the turning points of the particle's oscillations occur just below peaks and the particle spends a long time there, whereas near the trough the particle is moving fast and spends little time. Thus, the particle can never lower its wave field enough to escape to neighboring minima of its wave field and its motion is always confined between two consecutive peaks of its wave field. Moreover, since the amplitude of the particle-generated wave at each instant scales with $R$, the corresponding height of the wave field $Z$, its wave gradient $Y$ and the particle velocity $X$ also increase with $R$ resulting in an increase in size of the limit cycle with $R$ in phase space. However, since the particle can never escape its peaks, its motion remains bounded between two consecutive peaks and hence the amplitude of oscillations in particle position remains fixed.   %We note that these symmetry breaking transitions  can have toplogical implications for the attractors where links and knots may form~\citep{jackson_1990}. %These mergings have been studied topologically in the Lorenz system(Add ref.).

In this limit of large $R$, surprisingly, the system becomes integrable and reduces to solving second Painlev\'{e} transcendant that behaves asymptotically like elliptic functions~\citep{difflesslorenz2000}. There have been few studies that have explored this regime of DLE in detail~\citep{difflesslorenz2000,HUANG2003248,Ovsyannikov2023,Dong_2020} and we refer the interested reader to these papers.
%\begin{itemize}
%    \item In the limit the system is integrable and has an exact solution which can be stated and reference and then I can comment on the implications of this for the trajectory of the particle.
%    \item \textbf{Figures for this section:} (1) A plot showing evolution of limit cycle as R increases and if there are any other interesting feaures. (2) Multistability and basin of attraction where different types of runaway walkers co-exist with different average speed that corresponds to different topology of the limit cycle.
%\end{itemize}

\subsection{\label{sec: Multistab} Coexisting attractors and their basin of attraction}

%\textbf{(The WPE system's initial conditions are constrained by $x_d(0)=0$, $X(0)=X_0$ $Y(0)=R/X_0$ for $X(0)\neq0$ and $Z(0)=0$.)}

We have observed multistability in this system where phase-space attractors that correspond to different types of WPE motion coexist at the same $R$ value in the small, intermediate as well as large wave-amplitude regime. Some examples of coexisting attractors and their basin of attraction are shown in Figs.~\ref{Fig: multi}(a-d). Figure~\ref{Fig: multi}(a) and (b) show multistability in the small wave-amplitude regime for $R=0.06$ and $R=0.11$ respectively. For $R=0.06$, we observe the coexistence of self-trapped intermittent WPEs and runaway intermittent WPEs. The basin of attraction reveals a fractal structure with the basins of left and right moving runaway oscillating WPEs intricately intertwined in a sea of self-trapped oscillating WPEs. From the phase-space trajectories, we see that the two asymmetric limit cycles for runaway oscillating WPEs seem to have multiple links. For $R=0.11$, we find the coexistence of irregular intermittent WPEs and runaway intermittent WPEs. Here also we find an intricate fractal basin of attraction for the two runaway oscillating WPE attractors embedded in a sea of irregular WPE attractors. However the structure appears to be less complex as compared to the basin of attraction of $R=0.06$. This relatively low complexity is also reflected in the phase-space trajectory of the two asymmetric limit cycles for runaway WPEs which have one simple link. Figure~\ref{Fig: multi}(c) shows multistability for $R=0.22$ where irregular WPEs co-exist with self-trapped oscillating WPEs. Here the basin boundary is even more smooth and the two asymmetric runaway oscillating WPE attractors are replaced by a single symmetric self-trapped oscillating WPE attractor. Figure~\ref{Fig: multi}(d) shows multistability in the large wave-amplitude regime for $R=6$ where now two different kinds of runaway oscillating WPEs co-exist and again we find a fractal basin of attraction. Different types of coexisting runaway oscillating WPEs were also observed in the high-memory regime by \citet{Durey2020} using their Bessel wave form. % (refer to speed oscillation paper). By looking at the $X-Y$ projection of the attractor, one can see that there are no-crossings for the attractor in the middle panel while there is one crossing for the attractor in the right panel.%(\textbf{If i plot both the attractors for this solution then do they exhibit links? Are they different in the number of links?})    

From the viewpoint of a WPE, a typical experimental initial condition would correspond to the particle initially at rest with no wave-memory i.e. $(X(0),Y(0),Z(0))=(0,0,0)$. These lie right inside the fractal structure seen in these basin of attractions. Hence, if this regime can be realized in experiments with walking/superwalking droplets, then one might expect extreme sensitivity to initial conditions since typical initial conditions for WPEs in experiments are likely to fall in the fractal structure.

This aspect of multistability also enables an easy way to access different dynamical states from the viewpoint of an attractor-driven particle. By adding a small amount of noise in the internal state dynamics of the DLE for fixed $R$, the attractor-driven particle can transition from irregular motion to self-trapped motion or runaway motion as the internal state-space system switches between different types of phase-space attractors. Of course, this may also be achieved by tuning the control parameter $R$ in the appropriate regime. 

%\begin{itemize}
%    \item It will be good to have a bifurcation diagram of this system along with the multistability aspect. It seems that the system definitely shows multistability
%    \item For multistability, I can use the initial condition of the next simulation as the final state of the previous simulation to follow a particular branch of a solution
%    \item There definitely seems to be multistability as $R$ changes. I can potentially look at a bifurcation diagram and also basin of attraction for Multistability. Then where the motion is random, I can look at the statistics of various jumps and see how they change with the parameter $R$ in that range. Also, I can classify the intermittent hopping as pseudolaminar chaotic diffusion in a genralized sense and then look at the return maps of Plateau's and their distances. Would be good to connect this to laminar/pseudolaminar chaos
%    \item It seems like two topologically different asymmetric limit cycles exist that corresponds to different average speed of oscillating walkers. For example at $R=6$, two different oscillating walking states exist. We can look at the basin of attraction for such states in the initial condition space and classify them. This would be something new. We are now giving physical meaning to different phase-space strctures such as limit cycles and knots!! Maybe contact Thiago De Paiva Souza from Monash University who works with Jessica Purcell on Lorenz Links \& Knot Theory and Dynamical System
%\end{itemize}

\section{\label{sec: conclusion} Conclusions}

In this paper we have explored the rich dynamical behaviors of a classical active WPE in the limit of infinite wave-memory. We showed that the system reduced to one of the algebraically simplest chaotic systems, the diffusionless Lorenz equations (DLEs), with a single parameter $R$ representing the dimensionless wave-amplitude. The algebraic simplicity of ODEs is deceiving and the system exhibits rich dynamics and bifurcation structure which we have explored in the context of WPE motion and an attractor-driven particle. 

The rich dynamical behaviors observed for the WPE as a function of $R$ were classified into three distinct types: self-trapped oscillating WPE, runaway oscillating WPE and irregular WPE. In the small $R$ regime, these three types of dynamical behaviors were realized with intermittent dynamics where the WPE spends a long time in a stationary state while it is building/erasing its wave field and then swiftly takes a step of nearly half the wavelength. We linked this mechanism of intermittent motion to the corresponding dynamics taking place in the phase space of the system where each step in intermittent motion of WPE is related to an orbit around one wing of the corresponding phase-space attractor. \citet{Durey2020} in their infinite-dimensional integro-differential equation model for WPE dynamics with a Bessel wave form found similar trajectories. We find that our simple model that reduces to a system of three nonlinear ODE systems captures qualitative features of the more complete model. The bifurcation diagram in the small $R$ regime showed a self-similar period-doubling structure where all three types of motion exist. We explored chaotic aspects of irregular intermittent WPE where the return map of the step-length showed a low-dimensional structure and the trajectory showed similarities with pseudolaminar chaotic diffusion. We also explored the statistical properties by investigating the position distribution of particles and found wave-like statistics that persist for long times. Moreover, time-periodic fluctuations were observed in the position distribution at short and intermediate time scales. In the intermediate $R$ regime, the system exhibited mainly chaotic dynamics with the extent and complexity of the phase-space attractors increasing with $R$. In the large $R$ regime, a period halving bifurcation ceases chaos and one eventually gets symmetric limit cycles corresponding to self-trapped oscillations with the size of the limit cycle in phase-space increasing with increasing $R$ but the particle motion confined between two consecutive peaks of its wave field. We also showed multistability in the system where different types of motion coexist at the same $R$ and they are intricately interwoven in the basin of attraction. 

The rich set of dynamical behaviors exhibited by the DLE also gives our attractor-driven particle a diverse array of features that are not typically observed in traditional active particles. The single parameter $R$ provides a convenient way to assign different dynamical states to the attractor-driven particles and the presence of multistability further enhances this richness and provides ways to access different dynamical states at the same $R$ value. When the DLE system exhibits chaos on a strange attractor with the complexity of the attractor varying with $R$, this provides a way to tune the transport properties of the attractor-driven particle. Moreover, the intermittent motion for small $R$, gives rich statistical features to the attractor-driven particle such as spatial and temporal oscillations. This specific example of attractor-driven particle explored in this paper shows the richness of the framework of attractor-driven matter~\citep{Valaniattractormatter2023}.

The ODE framework of our simple Lorenz-like system enables a detailed exploration of three-dimensional phase-space attractors and their bifurcations allowing us to link the dynamics and geometry in phase-space to the motion and trajectories of the WPE or the attractor-driven particle. Even these deceptively simple looking Lorenz-like systems exhibit a complex array of behaviors in phase-space that have not been completely uncovered and research is still in progress to understand the interplay between geometry, dynamics and topology~\citep{Osinga2002,Osinga2018,Doedel_2006,Creaser_2015,DOEDEL2011222,ToplogyLorenz2017,doi:10.1073/pnas.2205552120}. A comprehensive understanding of phase-space behaviors associated with the underlying attractors of WPE systems may lead to new perspectives in rationalizing quantum-like statistics in hydrodynamic quantum analogs of walking droplets, and also new advances in active particle modeling using attractor-driven particles.

%\begin{itemize}
%    \item Can relate it to a simple example of attractor driven particle
%    \item Can relate it to ultra-chaos maybe
%    \item Pseudo-lam chaos connection
%    \item We hint at a link between the wave-like statistics in pilot-wave dynamics, to the phase-space dynamics of the underlying attractor and also to possibly of topoligical properties of the attractors. THis warrants further investigation.
%\end{itemize}

\begin{acknowledgments}
I would like to thank David Paganin for helpful comments and discussions. R.V. was supported by Australian Research Council (ARC) Discovery Project DP200100834 during the course of the work. Some of the numerical results were computed using supercomputing resources provided by the Phoenix HPC service at the University of Adelaide.
\end{acknowledgments}

\section*{Data Availability Statement}

The data that support the findings of this study are available from the corresponding author upon reasonable request.

\appendix

\section{Equilibrium points and linear stability of the wave-particle entity at finite and infinite memory}\label{sec: equilibrium}

The equilibrium states of the wave-particle entity for finite memory i.e. finite $\tau$ can be obtained by finding critical points of the system of ODEs in Eqs.~\eqref{lorenz}. If all the four ODEs are considered then one gets the stationary state $(x_d,X,Y,Z)=(x_0,0,0,0)$ as the equilibrium solution. To obtain steady walking states we can find critical points of the last three equations in \eqref{lorenz} i.e.
\begin{align*}
       \dot{X}&=Y-X, \\
    \dot{Y}&=-\frac{Y}{\tau}+XZ, \\
    \dot{Z}&=R-\frac{Z}{\tau}-XY. 
\end{align*}
This gives critical points
\begin{align*}
    X&=Y=\pm\sqrt{R-\frac{1}{\tau^2}},\\ 
    Z&=\frac{1}{\tau}.
\end{align*}
These correspond to the particle moving steadily to the right or left respectively. To check if this solution is consistent with the full system in Eqs.~\eqref{lorenz} and \eqref{eq: Y Z}, we substitute the right walking solution $X=\sqrt{R-\frac{1}{\tau^2}}$ in the first equation in \eqref{lorenz} giving (assuming $x_d(0)=0$ without loss of generality)
\begin{equation*}
    x_d(t) =  \sqrt{R-\frac{1}{\tau^2}} t.
\end{equation*}
Substituting this in Eq.~\eqref{eq: Y Z} we get,
\begin{align*}
    Y &= R \int_{-\infty}^{t} \sin\left( \sqrt{R-\frac{1}{\tau^2}} (t-s)\right)\,\text{e}^{-\frac{(t-s)}{{\tau}}}\,\text{d}s.
\end{align*}
By making a change of variables $t-s=z$, one gets,
\begin{align*}
    Y &= R \int_{0}^{\infty} \sin\left( \sqrt{R-\frac{1}{\tau^2}} z \right)\,\text{e}^{-\frac{z}{{\tau}}}\,\text{d}z,
\end{align*}
which results in 
\begin{align*}
    Y = R \left( \frac{\sqrt{R-\frac{1}{\tau^2}} }{R}\right) = \sqrt{R-\frac{1}{\tau^2}},
\end{align*}
given that $R\neq0$ and $1/\tau >0$. Similarly, for $Z(t)$ one gets,
\begin{align*}
    Z &= R \int_{0}^{\infty} \cos\left( \sqrt{R-\frac{1}{\tau^2}} z \right)\,\text{e}^{-\frac{(z)}{{\tau}}}\,\text{d}z,
\end{align*}
which results in
\begin{align*}
    Z = R \left(\frac{1/\tau}{R}\right) = 1/\tau,
\end{align*}
given that $R\neq0$ and $1/\tau >0$. Hence these solutions are consistent with the ones obtained ODEs.

Now, in the limit $\tau \rightarrow \infty$ we have $1/\tau \rightarrow 0^{+}$ so these solutions are still valid and we obtain the equilibrium solutions:
   \begin{align*}
    X&=Y=\pm\sqrt{R},\\ 
    Z&=0. 
\end{align*}

For the system at finite memory, applying a small perturbation to this equilibrium state $(X,Y,Z)=(\pm\sqrt{R-\frac{1}{\tau^2}},\pm\sqrt{R-\frac{1}{\tau^2}},0)+\epsilon(X_1,Y_1,Z_1)$, where $\epsilon>0$ is a small perturbation parameter, results in the following linear system that governs the leading order evolution of perturbations:
\begin{gather*}
 \begin{bmatrix} 
 \dot{X}_1 \\
 \dot{Y}_1 \\
 \dot{Z}_1 
 \end{bmatrix}
 =
  \begin{bmatrix}
-1 & 1 & 0 \\
\frac{1}{\tau} & -\frac{1}{\tau} & \pm\sqrt{R-\frac{1}{\tau^2}} \\
\mp\sqrt{R-\frac{1}{\tau^2}} & \mp\sqrt{R-\frac{1}{\tau^2}} & -\frac{1}{\tau}
 \end{bmatrix}
  \begin{bmatrix}
  {X}_1 \\
 {Y}_1 \\
 {Z}_1 
 \end{bmatrix}.
\end{gather*}
The linear stability is determined by the eigenvalues of the right-hand-side matrix. This results in the following characteristic polynomial equation to be solved for the eigenvalues $\lambda$ which determines the growth rate of perturbations:
$$\lambda^3 + \left(\frac{2}{\tau}+1\right)\lambda^2 + \left(R+\frac{1}{\tau}\right)\lambda + 2R-\frac{2}{\tau^2} =0.$$
Since $R-\frac{1}{\tau^2}>0$ for the steady walking state, by using Descartes' sign rule we we either have (i) one negative real eigenvalue and a complex conjugate pair or (ii) three negative real eigenvalues. We can get further clarity by finding the discriminant of this cubic eigenvalue equation which gives 
$$\Delta=4M^3-T^2M^2-18DTM+2TD^2+4DT^3,$$
with
$$T=-\left(1+\frac{2}{\tau}\right),$$
$$M=R+\frac{1}{\tau},$$
$$D=-2\left(R-\frac{1}{\tau^2}\right).$$
%\textit{If I can't get an explicit expression here then maybe I can just plot this function for $\Delta$ in a reasonable parameter space and show that its always above/below the linear stability curve.}
By plotting $\Delta$ in the $\tau-R$ space we find that its always negative resulting in a complex conjugate pair of eigenvalues and thus we have one negative real eigenvalue and a complex conjugate pair. 

To find the stability boundary of the steady walking state, one needs to know when the real part of the complex conjugate eigenvalues changes its sign. By substituting $\lambda=i\omega$ in the eigenvalue equation we get two equations. The first one gives the stability boundary
\begin{equation}\label{eq: Rcric}
    R=\frac{\frac{1}{\tau}\left( 1+\frac{4}{\tau} \right)}{1-\frac{2}{\tau}},
\end{equation}
and the second one determines the frequency of small oscillations just above the stability boundary
\begin{equation}
 \omega = \sqrt{R+\frac{1}{\tau}}.
\end{equation}
Above the $R$ value as define in Eq.~\eqref{eq: Rcric}, the real part of the complex conjugate eigenvalue is positive and below this curve its negative. Now, in the infinite memory limit $\tau \rightarrow \infty$, the stability boundary separating positive and negative real parts of the complex conjugate eigenvalue, Eq.~\eqref{eq: Rcric}, approaches $R\rightarrow 0$, and hence in the infinite memory limit we always have the real part of the complex conjugate eigenvalue as positive for $R>0$.

%\nocite{*}
\bibliography{aipsamp}% Produces the bibliography via BibTeX.

\end{document}